\documentclass[draft]{agujournal2019}
\usepackage{url} 
\usepackage{lineno}
\usepackage[inline]{trackchanges} 
\usepackage{soul}
\usepackage{url} 
\usepackage{lineno}
\usepackage{soul}
\usepackage{amsmath}
\usepackage{float}

\usepackage[export]{adjustbox}
\usepackage{graphicx}
\usepackage[french,english]{babel}
\usepackage[T1]{fontenc}
\usepackage[utf8x]{inputenc}
\usepackage{multicol}

\newcommand{\CO}{CO$_2$ }
\newcommand{\Ls}{L$_s$}
\newcommand{\sigT}{3-$\sigma$ }
\newcommand{\sigO}{1-$\sigma$ }

\draftfalse

\journalname{Journal of Geophysical Research: Planets}

\begin{document}

%
%

\title{InSight Pressure Data Recalibration, and its Application to the Study of Long-Term Pressure Changes on Mars}

%
%




\authors{L.Lange\affil{1}, F.Forget\affil{1}, D.Banfield\affil{2}, M.Wolff\affil{3}, A.Spiga\affil{1,4}, E.Millour\affil{1}, D. Viúdez-Moreiras\affil{5}, A.Bierjon\affil{1}, S.Piqueux\affil{6},  C.Newman\affil{7}, J.Pla-García\affil{5,8}, W.B.Banerdt\affil{6}}


\affiliation{1}{Laboratoire de Météorologie Dynamique,Institut Pierre-Simon Laplace (LMD/IPSL), Sorbonne Université, Centre National de la Recherche Scientifique (CNRS), École Polytechnique, École Normale Supérieure (ENS), Paris, France}
\affiliation{2}{Cornell Center for Astrophysics and Planetary Science, Cornell University, Ithaca, NY, USA}
\affiliation{3}{Space Science Institute, Boulder, CO, USA}
\affiliation{4}{Institut Universitaire de France, Paris, France}
\affiliation{5}{Centro de Astrobiología (CSIC-INTA) and National Institute for Aerospace Technology (INTA), Madrid, Spain}
\affiliation{6}{Jet Propulsion Laboratory, California Institute of Technology, Pasadena, CA 91109, USA}

\affiliation{7}{Aeolis Research, Unit 5, Chandler, AZ, USA}
\affiliation{8}{Southwest Research Institute, Boulder, CO, USA}





\correspondingauthor{Lucas Lange}{lucas.lange@lmd.ipsl.fr}




\begin{keypoints}

\item We propose a recalibration of InSight pressure data to correct an unexpected  sensitivity  to the sensor temperature;
\item A comparison between the InSight  and Viking 1 pressure data does not show secular changes in the global mass of the atmosphere;
\item This comparison also supports the absence of long-term variability in the dynamics of seasonal cap formation and sublimation.
\end{keypoints}

%
%

%
%


\begin{abstract}

Observations of the South Polar Residual Cap suggest a possible erosion of the cap, leading to an increase of the global mass of the atmosphere. We test this assumption by making the first comparison between Viking 1 and InSight surface pressure data, which were recorded ~40 years apart. Such a comparison also allows us to determine changes in the dynamics of the seasonal ice caps between these two periods. To do so, we first had to recalibrate the InSight pressure data because of their unexpected sensitivity to the sensor temperature. Then, we had to design a procedure to compare distant pressure measurements. We propose two surface pressure interpolation methods at the local and global scale to do the comparison. The comparison of Viking and InSight seasonal surface pressure variations does not show major changes in the \CO cycle. Such conclusions are also supported by an analysis of Mars Science Laboratory (MSL) pressure data. Further comparisons with images of the south seasonal cap taken by the Viking 2 orbiter and MARCI camera do not display significant changes in the dynamics of this cap over a ~40 year period. Only a possible larger extension of the North Cap after the global storm of MY 34 is observed, but the physical mechanisms behind this anomaly are not well determined. Finally, the first comparison of MSL and InSight pressure data suggests a pressure deficit at Gale crater during southern summer, possibly resulting from a large presence of dust suspended within the crater. 

\end{abstract}

\section*{Plain Language Summary}
Observations of the permanent \CO ice cap at the south pole of Mars in the 2000s suggested that the cap was eroding, possibly releasing a significant amount of \CO into the atmosphere. To test this hypothesis, we compare surface pressures recorded by Viking in the 1970s and those recorded by InSight in 2018-2021 to confirm or refute the suspected increase of the atmospheric mass. After establishing our comparison method, we correct  the influence of the sensor temperature on the InSight pressure data, which was discovered during our investigation. Comparison of the pressure data, as well as images of the seasonal caps taken by orbiters, do not reveal any change in the atmospheric mass or the dynamics of the seasonal ice caps that develop during the martian year. These conclusions are reinforced by re-exploiting the pressure data recorded by the Curiosity rover. Only small interannual changes are observed, potentially related to the effect of the dust storms that happened on Mars between 2016 and 2018. Finally, we report a possible pressure deficit at MSL's location during southern hemisphere summer, potentially explained by the unusual presence of dust in the crater air.

%
%

%


%
%
%
%

\section{Introduction}
\label{sec:Intro}

The retreat of the Southern Seasonal Polar Cap (SSPC) during local summer leaves a residual perennial deposit mainly composed of  CO$_2$ ice \cite{KIEFFER1972}. This deposit, known as the South Polar Residual Cap (SPRC), is one of the  CO$_2$ reservoirs that can significantly affect the atmospheric mass through sublimation or deposition \cite[]{Leighton1966}. The stability of this reservoir over time is a long-standing debate in Martian climate science. While \citeA{BLACKBURN2010780} predicted the disappearance of the SPRC within a few years, \citeA{Piqueux2008} reported limited changes in the extent and ice-covered area of the SPRC since the Mariner 9 mission in 1972 and telescopic observations in the twentieth century. However, \citeA{Piqueux2008} were unable to retrieve a mass balance of the cap. Other monitoring of the SPRC surface since the Mariner 9 and Viking era led to mass balance estimates suggesting either an erosion of the SPRC \cite[]{Malin2001, THOMAS2009352, BLACKBURN2010780, THOMAS2013923, THOMAS2016118} or a possible ice accretion \cite[]{THOMAS2016118}. 
This presumed erosion or accretion of the SPRC open the possibility of  secular pressure changes on Mars: if the SPRC loses \CO ice year after year, the sublimated \CO ice goes into the atmosphere, increasing its global mass and thus also the global mean surface pressure on Mars \cite[]{Malin2001, Kahre2010, Haberle2010, THOMAS2016118}. Conversely, the deposition of \CO ice on the cap would decrease the atmospheric mass, and thus the atmospheric pressure. Observations from \citeA{Malin2001} suggested that the CO$_2$ ice thickness decreases by nearly 0.4 m per Mars Year (MY), which means an increase of surface pressure of almost +13 Pa per Martian Decade (MD). \citeA{BLACKBURN2010780} also estimated a possible increase of surface pressure between  +0.5 and +13 Pa/MD. A recent study by \citeA{THOMAS2016118} qualified the two previous estimations by reporting a much smaller variation of SPRC mass balance, with a possible variation of surface pressure between -2.3 Pa/MD and +1.6 Pa/MD. 

In addition to this possible secular change in atmospheric mass, we can investigate the possibility of an unknown mechanism that would change the dynamics of formation and sublimation of the \CO seasonal caps. Thermal infrared observations \cite{PIQUEUX2015} or cap albedo monitoring \cite{CALVIN2015North, CALVIN2017South} have already reported interannual variability in the formation and recessions of the seasonal caps. However, these studies are limited to a few years of observations, thus preventing a generalization of a possible secular change in the dynamics of seasonal ice caps.

A direct way to assess long-term pressure changes on Mars consists of comparing surface pressure measurements separated by several martian decades. By this method, we can check if the atmospheric mass has changed,  and study possible variability in the dynamic of the seasonal ice caps. Such a comparison must be done carefully, however, because of the influence of orography and meteorological variability on the annual surface pressure cycle  \cite[]{Hourdin1993, Hourdin1995}. The comparison of pressure measurements made by Viking  between 1976 and 1982 and those by  Phoenix in 1997, after being corrected for both topography differences and atmospheric dynamics simulated by a Global Circulation Model (GCM), showed a possible  10~Pa rise of surface pressure \cite{Haberle2010} which corresponds to 5~Pa/MD. However, the combined uncertainties in both the measurements and the interpolation methodology were not sufficiently accurate to draw any conclusions about a secular pressure change. More recently, the comparison between Mars Science Laboratory pressure measurements, which have been recorded since 2012, and Viking measurements did not show significant changes in  surface pressure  \cite[]{Haberle2014}. However, these conclusions are limited by the sensitivity to the hydrostatic adjustment of surface pressure as the rover is climbing Mount Sharp in Gale Crater \cite[]{Haberle2014, RICHARDSON2018132}, and the sensitivity of the atmospheric dynamics at Gale Crater that have to be resolved by a mesoscale model \cite[]{PLAGARCIA2016103, RAFKIN2016114}. Hence, the analysis of the first available surface pressure data neither confirmed nor denied any long-term pressure changes.

In 2018, the InSight mission deployed a geophysical and meteorological observatory, including a pressure sensor, at the surface of Mars \cite[]{Banfield2018, Banerdt2020}. The instruments are deployed on a static lander at Elysium Planitia, a relatively flat area located at 4.5\textdegree~N, 135.6\textdegree~E \cite[]{Golombek2020}, thus reducing the sensitivity of surface pressure measurements to both hydrostatic adjustment and atmospheric dynamics. Pre-flight calibration and tests suggested that the performances of the sensor were good enough to detect changes in the atmospheric mass and CO$_2$ cycle \cite[]{Spiga2018}. We thus propose in this paper to compare the InSight pressure data with the Viking pressure data to assess the possibility of long-term pressure changes over two Mars decades.

We  present in section \ref{sec:MethodInterp} the methodology of the pressure interpolation that will lead our comparison between Viking and InSight data. A closer look at the InSight data reveals a calibration problem due to a sensor temperature sensitivity. We  propose an empirical recalibration and test the reliability of this correction in section \ref{sec:RecalibrationGlobal}. The comparison between the InSight corrected pressure data and Viking surface pressure measurements is then presented in section \ref{sec:Results} to check for a possible secular increase or decrease of atmospheric mass. Long-term variations in the dynamics of the seasonal ice caps between the 1970s and 2018 are also investigated using pressure data and satellite images from the Viking and InSight eras, respectively. We then extend the scope of this study by also exploiting the Phoenix and MSL measurements to detect any evolution of the atmospheric mass in section \ref{sec:Discussion}. We also look at the possible influence of interannual variability of the seasonal cap due to the dust cycle. The main conclusions from our investigation are presented in section \ref{sec:conclusion}.

\section{Methodology of Pressure Interpolation \label{sec:MethodInterp}}

 The interpolation of the Viking pressure to the InSight landing site requires taking into account  planetary-scale atmospheric dynamics
that affect the surface pressure \cite[]{Hourdin1993, Hourdin1995}. Even local interpolation between two close points must include the influence of local weather phenomena, like slope winds. Hence, interpolating pressure cannot be limited to integrating the hydrostatic equation to correct for altitude differences. To take into account the impact of atmospheric dynamics at all scales, we propose two high-accuracy interpolation methods: one on a local scale (typically within a crater, a slope, etc.), and one on a regional-to-global scale.

\subsection{Local pressure interpolation}
\label{ssec:InterpLoc}
We consider here a local situation in which two points are close enough so that large-scale dynamic pressure gradients related to the global atmospheric circulation and regional flows can be neglected.  Let us consider two points A and B located at different altitudes (Figure \ref{fig:ScaleH}a). Since 
these two points are close, the main factor that impacts the difference in the absolute pressure is altitude, thus we could assume hydrostatic equilibrium and recast pressure  ($P_B$)  at  point B to the altitude at point A ($P_{B\rightarrow A}$) with:

\begin{equation}
    P_{B\rightarrow A} = P_{B}e^{-\frac{z_{A} - z_{B}}{H}}
    \label{Eq:Hydrostatic}
\end{equation}
where $z$ corresponds to the altitude of each point, $H$ is the scale height expressed as:

\begin{equation}
   H = \frac{R T}{\mu g}
    \label{Eq:H}
\end{equation}

\noindent with $R = 8.3145$ J~kg$^{−1}$~mol$^{-1}$ the molar gas constant, $T$ is the mean atmospheric temperature between A and B weighted by vertical pressure field (in Kelvin), $\mu = 43.34 \times 10^{-3}$ kg~mol$^{-1}$ the mean molecular weight of Mars atmosphere and $g = 3.72$ m~s$^{-2}$ Mars' surface gravity.

On terrains with an uneven topography,  local circulations, like slope winds, can appear as a consequence of  temperature gradients. Hence, the choice of a scale height $H$, and thus the temperature to take into account in Eq. \ref{Eq:H}, is important to consider the major effect of these local circulations \cite[]{Spiga2007, Forget2007}: the temperature choice in $H$ will indicate which path should be used to integrate the hydrostatic equation (red and green lines in Figure \ref{fig:ScaleH}a).  Such effects are very important for the Mars Science Laboratory mission for instance. As the Curiosity rover moves in Gale Crater, with significant gains of elevation (several hundred meters), local circulations and slope winds \cite[]{PLAGARCIA2016103, RAFKIN2016114, RICHARDSON2018132} also contribute to the absolute pressure recorded by the rover. \citeA{Forget2007, Spiga2007} suggested using the temperature at an  altitude of 1 km above the surface in Eq. \ref{Eq:H} to take into account the effect of slope winds at the GCM scale, while \citeA{ORDONEZETXEBERRIA2019591} used the air temperature at an altitude of 2 m when using MSL pressure data. \citeA{Haberle2014} also questioned the choice of the scale height $H$  that has to be used when exploiting MSL data. Their study of the sensitivity of pressure data to this scale height shows that, with extreme temperature scenarios, the absolute pressure can be influenced by several Pascals. However, they never determine which scale height is the right one to use with these data.

Thus, we investigate here the scale height that better matches MSL observations, and quantify the errors made during the interpolation of the surface pressure, using the example of Gale Crater. To do so, simulations of Gale Crater with the LMD mesoscale model \cite{Spiga2009} were performed. The domain for the simulations ranges  from 22\textdegree~S to 30\textdegree~N and 108\textdegree~E to 163\textdegree~E, with a spatial resolution of 0.16\textdegree~, including thus the InSight landing site, Gale Crater and its circulation.

We take the diurnal cycle of surface pressure simulated at several seasons at grid points at the bottom (B) and the rim of Gale Crater (A) ($\Delta z = 1725$m, in the axis of MSL trajectory). We then interpolate the pressure at point B ($P_B$) to location A using several altitudes for the temperature above the point B to compute $H$. We then compute the relative error between the exact modeled pressure at A ($P_A$), and the interpolated pressure from B to A ($ P_{B\rightarrow A}$). Results are presented in Figure  \ref{fig:ScaleH}b. They show that choosing the temperature at an altitude between 500 m and 2 km above the surface is better to take into account the effect of local dynamics on the pressure interpolation as it minimizes the relative difference between $P_A$ and $ P_{B\rightarrow A}$.  In the following, we choose to compute the scale height $H$  by using the temperature at an altitude of 1 km. When interpolating actual measurements, this temperature is not available from observations and instead has to be estimated using an atmospheric model. The main uncertainty in this calculation results from the sensitivity of temperatures to the dust opacity, which is not perfectly known. To check the sensitivity of the interpolation to these weather conditions and thus to an error in $T$, we use the GCM runs to quantify the impact of the dust opacity using three dust scenarios as input:
\begin{itemize}
    \item A climatology (\textit{clim}) scenario, derived by averaging the available observations of dust from MY 24, 25, 26, 28, 29, 30, and 31 outside the global dust storm period \cite{MONTABONE201565}. This scenario represents a nominal dust scenario in the absence of major dust storms.
    \item A \textit{cold} scenario which corresponds to an extremely clear atmosphere. At a given date and location, the dust opacity is set to be the minimum observed over Mars years 24-31, further decreased by 50\%. 
    \item A \textit{warm} scenario which corresponds to “dusty atmosphere” conditions, outside of global dust storms. The dust opacity at a given location and date is set to the maximum observed over seven Mars years (MY 24-MY 31), excluding the periods of the MY 25 and MY 28 global dust storms, further increased by 50\%.
\end{itemize}
These scenarios are used in the Mars Climate Database (MCD, \citeA{Millour2018}), and frame well the different temperature observations made by several spacecrafts at a \sigT level \cite{Millour2018}. Using these scenarios, our simulations show that the temperature of the air at an altitude of 1 km can vary by several kelvins. We consider the worst-case scenario, assuming that the \textit{cold} scenario decreases the temperature by 10 K compared to the \textit{clim} scenario; and the \textit{warm} scenario increases the temperature by 10 K  (simulations report a maximum of $\pm $ 8 K in terms of anomaly, and we add a 2 K margin). The relative errors made in the interpolation by using these temperature deviations instead of nominal temperatures are presented in Figure \ref{fig:ScaleH}c. This sensitivity study shows that the relative \sigT accuracy of this interpolation method  is almost 1\%, and is thus acceptable to exploit MSL pressure data. In summary, we found that an accurate  way to interpolate surface pressure from a point B to a point A at a local scale consists of using Eq. \ref{Eq:Hydrostatic} with the scale height computed using the temperature at an altitude of 1 km above point B (Eq. \ref{Eq:H}).

\subsection{Large-scale pressure interpolation}
\label{ssec:GlobalInterp}

At the planetary scale, in addition to the hydrostatic adjustment and local dynamic effects, we must take into account large-scale dynamic pressure gradients in the interpolation \cite{Hourdin1993, Hourdin1995}. Hence, interpolating Viking pressure to InSight cannot be done by using Eq. \ref{Eq:Hydrostatic} alone, as it does not consider these gradients.

To account for these atmospheric large-scale dynamic components, we use a method based on the LMD GCM  \cite[]{Hourdin1993, Forget1999}. Practically, the interpolation of Viking pressure data to obtain the pressure at any location on Mars consists of determining the spatial variation of surface pressure in the GCM, with interpolation from the coarse GCM topography grid (5.625\textdegree~ in longitude, 3.75\textdegree~ in latitude) to the high-resolution MOLA grid (32 pixels per degree), plus a correction to perfectly match the seasonal variations at the Viking 1 site. 

Thus, the interpolation of Viking 1 surface pressure at any location, $P_s$, is done with \cite[]{Forget2007,Spiga2007}:

\begin{equation}
    P_s =  <P_\text{Viking}>  \frac{P_\text{GCM}}{<P_\text{GCM,Viking }>}e^{-\frac{z-z_\text{GCM}}{H}}
    \label{Interp}
\end{equation}

where $P_\text{GCM}$ is the pressure predicted by the GCM at the site we want to interpolate to,  $<P_\text{Viking }>$ corresponds to the pressure records of Viking averaged over 15 sols to remove any weather variations (thermal tides and transient waves),  $<P_\text{GCM, Viking }>$ is the surface pressure predicted by the GCM at the location of Viking 1 and also smoothed over 15 sols. $z-z_\text{GCM}$ is the difference between the MOLA altitude and the altitude defined with the interpolation of the GCM topography grid at the location we consider, and $H$ corresponds to the scale height computed with Eq. \ref{Eq:H}. The same procedure as the one used in section \ref{ssec:InterpLoc}, using the GCM pressure field binned every 2 hours, again shows that we must consider the temperature at an altitude of 1 km above the surface. In this expression (Eq. \ref{Interp}), $<P_\text{Viking}>$ is the pressure we want to interpolate, $\frac{P_\text{GCM}}{<P_\text{GCM, Viking }>}$ is the correction of atmospheric dynamics  induced by the pressure gradients, and $e^{-\frac{z-z_\text{GCM}}{H}}$ is a hydrostatic correction, taking into account the effect of local dynamics.

We use in  this study the Viking 1 surface pressure data rather than Viking 2 data. Viking 1 data are indeed more complete after removing the  measurements made during dust storms, less sensitive to baroclinic activity \cite[]{Ryan1981, Tillman1989, Tillman1993}, and closer to InSight than Viking 2 \cite[]{Morris1980, Golombek2020}, thus limiting the sensitivity to errors in the correction of the dynamics of the atmosphere.

The uncertainty of the interpolation depends on two independent uncertainties: one linked to the Viking 1 pressure measurements and one to the weather-induced uncertainty. On the one hand, pre-flight tests showed that the precision of the Viking pressure sensors was better than $\pm 0.2$\% of the readings, plus a term due to a temperature dependency of nearly 0.18\% \cite{Seiff1977, Mitchell1977}. Consequently, the precision of the pressure measurements was $\pm 3.4$ Pa for Viking 1. Such errors in the precision can be mitigated, however, as we are using a pressure signal averaged over 15 sols. Assuming that this precision error on a single measurement can be modeled by white noise with a \sigT of 3.4 Pa, we can reduce the uncertainty on the diurnal average pressure value by a factor $\sqrt{N}$ where $N$ is the number of measurements used for the diurnal or 15 Sols average. Typically, 200 measurements per sol are used to compute the diurnal average  \cite[]{Barnes1980}. Therefore, by using a 15 Sols averaged surface pressure in Eq. \ref{Interp}, the relative sensitivity to the uncertainty of Viking measurements is limited to 0.06 Pa, and is thus completely negligible in the following. 

On the other hand, Viking measurements are also impacted by systematic errors due to the instrumental drift, the 8-bit telemetry resolution, and the uncertainty on the zero level of the pressure sensor's output voltage. Based on the apparent stability of the sensor because of the great repeatability of the pressure data from one year to another outside dust storm periods \cite{Hess1980, Tillman1993}, the instrumental drift had been estimated to be only -0.1 $\pm$1 Pa per Earth year and will be neglected in the following. The error due to the 8-bit telemetry resolution \cite{Seiff1976, Tillman1993} yields an uncertainty of 8.8 Pa in the absolute pressure level for one single measurement, even if the sensor has a theoretical resolution of nearly 1~Pa  \cite[]{Hess1976, Seiff1977}. Assuming that this uncertainty on a single measurement can be modeled by white noise with a \sigT of 8.8 Pa, and using a 15 sols averaged surface pressure, this uncertainty is reduced to 0.16 Pa and will also be neglected in the following. 
The last systematic error related to Viking data is due to the uncertainty on the zero level of the pressure sensor's output voltage. This was determined by readings taken just before atmospheric entry. The resolution uncertainty in these zero readings causes an uncertainty of up to 8.8 Pa in the absolute pressure level \cite{Seiff1977, Kahanpaa2021}. \citeA{Hess1980} proposed adding 4.4 Pa to each measurement as it should be the best estimate of the true pressure measurements, reducing the absolute error by half. However, it remains unclear if this adjustment has been implemented in the Planetary Data System (PDS) where data are available. We will thus consider in the rest of the study that the absolute error on Viking 1 pressure measurement is $ \Delta P_\text{Viking} =$ 8.8 Pa at a \sigT.

\begin{figure}[H]
    \includegraphics[scale=0.4]{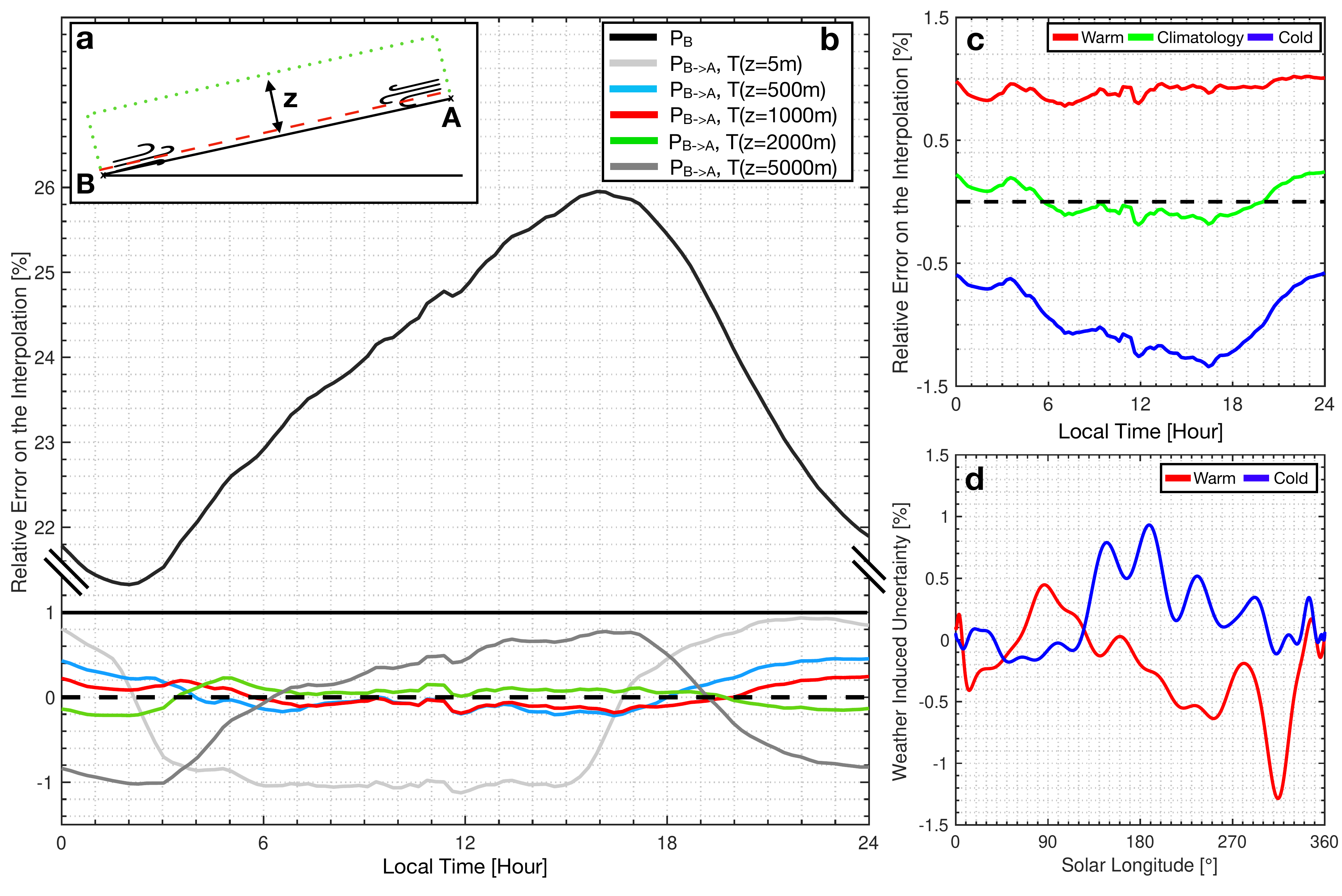}
    \caption{a) Schematics of the problem of interpolation with slope winds between the bottom of Gale Crater (point B) and the rim of the crater (point A). Colored dots illustrate the different paths that can be taken to integrate the hydrostatic equation. b) Relative error of the interpolated pressure from point B to point A and the exact pressure at A. The black curve is the relative error when point B is not interpolated to point A, while colored curves are for the relative errors when using different altitudes for the temperature. c) Relative error on the local interpolation when using the temperature at 1 km above the surface when considering several kinds of weather scenarios.  d) Weather-induced uncertainty of the Viking surface  pressure interpolated to InSight landing site computed with extreme dust scenarios when compared to \textit{clim} dust scenario (red and blue curves).}
    \label{fig:ScaleH}
\end{figure}

The second uncertainty in the interpolation is the influence of weather conditions, which impacts $T$ and thus $H$ in Eq. \ref{Interp} as well as the pressure predicted in the GCM. To study the impact of these conditions on the GCM output, we compute the interpolation of Viking 1 pressure to the InSight landing site by using the three dust scenarios described above, as they bracket well the possible states of the atmosphere \cite{Millour2018}. We then compute the weather-induced uncertainty, defined as the relative difference between the pressure at InSight's location derived with the extreme dust scenarios (\textit{cold} and \textit{warm}) and the \textit{clim} dust scenario (Figure \ref{fig:ScaleH}d).  Figure  \ref{fig:ScaleH}d underlines that this weather-induced uncertainty is generally limited to 1\% of the absolute pressure at \sigT. We set this uncertainty as 1\% of the  mean annual pressure of $700$ Pa at InSight's landing site (Figure \ref{fig:InsDiuAvgRaw}), i.e., by $\Delta P_\text{weather} = 7$ Pa at \sigT.  It should be noted that we use dust scenarios derived from Mars Climate Sounder (MCS, \citeA{McCleese2007}) observations from MY 29 to MY 35 \cite[]{MONTABONE201565, Montabone2020} for our comparisons.  The weather-induced uncertainty is therefore much lower as these accurate observational scenarios help to compute $T$, and thus $H$, in Eq. \ref{Interp} precisely.

Combining the independent uncertainty of Viking measurements $ \Delta P_\text{Viking} $ and the weather-induced uncertainty $\Delta P_\text{weather} $ yields an uncertainty of the interpolation of nearly 11 Pa at \sigT. Such a threshold is at the limit of the lowest predictions of atmospheric mass variations possibly indicated by cap studies \cite{THOMAS2016118}, but well below the first estimates made at the beginning of the 2000s \cite{Malin2001, BLACKBURN2010780}.

\section{Recalibration of InSight Pressure Data\label{sec:RecalibrationGlobal}}

\begin{figure}[H]
    
    \includegraphics[scale=0.45]{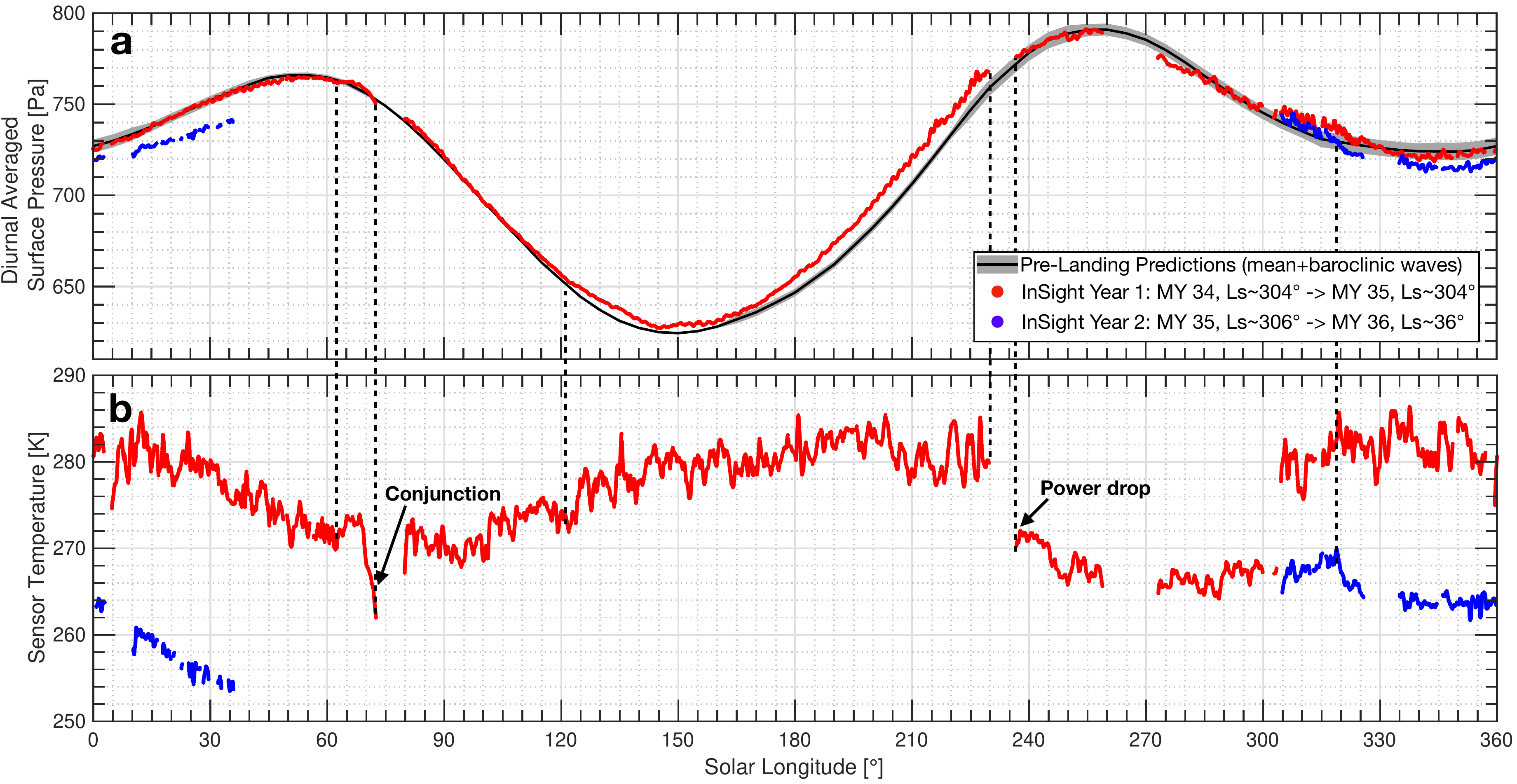}
    \caption{a) Diurnal averaged surface pressure computed from the 20 Hz data acquired during the two years of the mission (red and blue), with pre-landing surface pressure predictions (black curve) and baroclinic waves amplitudes (grey filled area)  from \citeA{Spiga2018}. b) Diurnal averaged temperature of the pressure sensor. Red dots are for the first year of the mission, while blue dots represent the measurements taken during the second year of the mission. Dashed black lines highlight the significant correlations between the sensor temperature and the pressure measurements.}
    \label{fig:InsDiuAvgRaw}
\end{figure}

The InSight pressure sensor is located on the lander deck at a height of approximately 1.2 m, with a sampling rate of up to 20 Hz and a noise level lower than 10 mPa, which theoretically represents an unprecedented quality compared to the different pressure sensors that have operated on the surface of Mars \cite[]{Banfield2018,Banfield2020,Spiga2018}. These data are calibrated by using output voltage and sensor temperature channels. We use in this study the 20 Hz data and average them with a 50s window to remove any effects of high-frequency pressure events (e.g.,\citeA{Chatain2021,Spiga2021}). We then compute the diurnal average of these signals. To do so, we interpolate the data from previous and following sols to complete the diurnal cycle when there are small gaps (typically of a few hours) in the data. We then interpolate the measurements onto a regular temporal grid containing 100 points per sol. From these interpolated points, we compute the diurnal average. The diurnally averaged surface pressure obtained for the entire dataset ($\sim$1.25 MY) is presented in Figure \ref{fig:InsDiuAvgRaw}a.

\subsection{Sensor temperature sensitivity of the InSight pressure data}
\label{ssec:ThermalSensitivity}

A direct comparison between the pressure measurements made one year apart during the first and second Martian years of the InSight mission shows a large difference (Fig. ??). This cannot be explained by the instrumental drift reported in \citeA{Banfield2018,Spiga2018} or by any likely major meteorological event as no significant long-lived global events have been observed. This difference is also not observed by MSL pressure measurements, thus questioning the reliability of the InSight pressure measurements. Furthermore, the divergence between the measurements made two years apart seemed to increase toward the end of the mission, when the power allocated to the pressure sensor was very low because of the accumulation of dust on the solar panels, leading to a decrease of the sensor temperature. A close comparison of the pressure measurements and sensor temperature (Figure \ref{fig:InsDiuAvgRaw}b) reveals that the pressure measurements are very likely to be affected by some drops or rises in the sensor temperature. An illustration of this correlation is at L$_s \sim$ 72\textdegree~, just before conjunction. Sensors were powered off, after the warm Electronics Box had cooled off, and a nonphysical drop occurred in the pressure measurements. This correlation questions the reliability of the calibration of the absolute pressure data to the sensor temperature. The pre-flight calibration of the pressure sensor with temperature may not be as accurate as expected, possibly because of the existence of temperature gradients within the instrument under real martian conditions. Such an effect had already been identified as responsible for pressure measurement errors on Phoenix \cite{Taylor2010}. It is important to note that most of the scientific results obtained from the pressure data are not impacted by this calibration problem. These works (see for instance \citeA{Banerdt2020,Spiga2021,Chatain2021}) use relative pressure variations and not absolute measurements, and at high frequencies. At these frequencies, i.e., for timescales of the order of a sol, or less, the fluctuations of the sensor temperature are negligible. The calibration problem detected is thus nullified when using relative variations of measured pressures, and therefore does not bias the scientific results obtained. We propose in section \ref{ssec:Recalibration} to correct this thermal effect using MSL pressure data.

\subsection{Mars Science Laboratory pressure data}
\label{ssec:MSLDATA}

In the absence of major meteorological events, we can expect limited interannual variations between the pressure measured during the first and the second year of the InSight mission (see legend of Figure \ref{fig:InsDiuAvgRaw} for definitions). \citeA{delaTorreJuarez2019} reported a strong interannual variability of the pressure data at the end of MY 34 and the beginning of MY 35 compared to other years using MSL pressure data. Such a difference might be linked to the global dust storm of MY 34 that possibly had an impact on the extent of the NSPC. 

To take into account a possible interannual variability between InSight's first and second years of the mission, we use MSL Rover Environmental Monitoring Station (REMS) pressure data available in the PDS. The calibrated data extracted from the PDS extend from MY 31, at a solar longitude (\Ls, the Mars-Sun angle, measured from the Northern Hemisphere spring equinox where \Ls=0\textdegree~) \Ls $ \sim$ 150\textdegree~ to MY 36, \Ls $ \sim $21\textdegree~.  The REMS pressure sensor acquires data during the first five minutes of each Local Mean Solar Time (LMST) with additional hour-long acquisitions that cover a full diurnal about every 6 sols \cite[]{Gomez2014}. To take into account the vertical displacement of the rover on pressure measurements, we interpolate the pressure data from the position of the rover determined with the Ancillary Data Record (ADR) files to the MSL landing site using the method described in section \ref{ssec:InterpLoc}.  The air temperatures at an altitude of 1 km above the surface are computed with the MCD, using dust scenarios from \citeA{MONTABONE201565, Montabone2020} as inputs for the simulations.

\subsection{Recalibration of the pressure measurements}
\label{ssec:Recalibration}

We define $E(T(t))$ (in Pa) as the sensitivity of the pressure measurements with regards to the sensor temperature $T$. The corrected measured pressure $ P_\text{InSight,Corrected}$ at a time $t$ can be written as:

\begin{equation}
   P_\text{InSight,Corrected}(t) = P_\text{InSight,Measured}(t) + E(T(t))
   \label{defP}
\end{equation}

We average the pressure measured by MSL and InSight over 15 sols to eliminate the contribution of any dynamical component like thermal tides and baroclinic activity. These averaged pressure values are denoted $<P>$ in the following. As InSight and MSL are relatively close ($\sim$ 600~km), we assume that the correction of the large-scale atmospheric dynamics between the two sites can be neglected. Our simulations show indeed that this correction is limited to 1~Pa at \sigT thus is negligible.

During InSight year 1 ($Yr_1$~:  MY 34, \Ls $\sim $ 304\textdegree~ to  MY 35, \Ls $\sim$ 304\textdegree~) and  year 2 ($Yr_2$~: MY 35, \Ls $\sim$ 306\textdegree~ to  MY 36, \Ls $\sim$ 36\textdegree~) of the mission, we have at first order:

\begin{equation}
\begin{cases}  
<P_\text{InSight,Corrected}(t_{Yr_1})> ~=~ <P_\text{MSL}(t_{Yr_1})>e^{-\frac{\Delta z}{H(t_{Yr_1})}}
\\ 
<P_\text{InSight,Corrected}(t_{Yr_2})> ~=~ <P_\text{MSL}(t_{Yr_2})>e^{-\frac{\Delta z}{H(t_{Yr_2})}}
\end{cases}
\label{EqMSLINS}
\end{equation}

\noindent with $\Delta z$ the difference of altitude between the InSight and MSL landing site, and $H$ the scale height computed with the air temperature at an altitude of 1 km above the surface. GCM computations show that with  MY 34, 35 and \textit{clim} dust scenarios, we have to first order $e^{-\frac{\Delta z}{H(t_{Yr_1})}} \sim e^{-\frac{\Delta z}{H(t_{Yr_2})}}$. Thus Eq.  \ref{EqMSLINS}  leads to:

\begin{equation}
    \frac{<P_\text{InSight,Corrected}(t_{Yr_1})}{<P_\text{InSight,Corrected}(t_{Yr_2})>} = \frac{<P_\text{MSL}(t_{Yr_1})>}{<P_\text{MSL}(t_{Yr_2})>} ~=~ \beta
    \label{fracbeta}
\end{equation}

where $\beta$ is by definition the interannual variability between the two years of measurements. Hence, as we only use a ratio of pressures, the absolute pressure values measured by MSL do not impact the absolute values of InSight pressure measurements after being corrected, and thus do not introduce a bias in our comparison. The problem described by Eq.  \ref{fracbeta} can be transformed into the following optimization problem: 

\begin{multline}
  \label{PbOpt}
  {\rm Find \ } E{\rm  \ that \ minimizes:}  \\
 \mid \mid   <P_\text{InSight,Corrected}(t_{Yr_1})> ~-~ \beta<P_\text{InSight,Corrected}(t_{Yr_2})> \mid \mid  
\end{multline}

Introducing Eq.  \ref{defP} into \ref{fracbeta}  gives:

\begin{multline}
  \label{EqF}
 <P_\text{InSight,Measured}(t_{Yr_1})> - \beta<P_\text{InSight,Measured}(t_{Yr_2})>  \\
= \beta<E(T((t_{Yr_2}))> ~-~ <E(T((t_{Yr_1}))>
\end{multline}

We further assume that $E$ can be written as a polynomial function of the sensor temperature: 

\begin{equation}
    E(T(t)) = \sum_{k = 0}^n \alpha_k T(t)^k
    \label{offset}
\end{equation}

Introducing this into Eq. \ref{EqF} finally leads to an expression of the function that we want to minimize:

\begin{multline}
  \label{EqLMS}
    <P_\text{InSight,Measured}(t_{Yr_1})> ~-~ \beta<P_\text{InSight,Measured}(t_{Yr_2})> \\
    = \sum_{k = 0}^n  \alpha_k <\beta T(t_{Yr_2})^k  ~-~ T(t_{Yr_1})^k>
\end{multline}

This last equation represents a least-mean-square problem that can be solved numerically to determine the coefficients $\alpha_k$ of $E$ for a given degree $n$. However, the problem must be constrained to have a physical solution. The first term $\alpha_0$ is indeed poorly constrained as $\beta \sim $1. A close look at  Figure \ref{fig:InsDiuAvgRaw} reveals an unexpected increase of the uncorrected pressure at \Ls $\sim$ 63\textdegree~, and then a drop, both certainly resulting from a rise of temperature at $T$ = 270~K and followed by a decrease of temperature at $T$ = 275~K. Such observations are also found at \Ls = 130\textdegree~ and 235\textdegree~, suggesting a change of behavior of the sensor  temperature sensitivity, i.e., a change in the sign  $E(T)$ close to $T$ = 0\textdegree~C. Furthermore, the fits used for the calibration of the pressure sensor are  good near $T $= 0\textdegree~C. Hence we simply assume that:

\begin{equation}
    E(T = 0^\circ C ) = 0~Pa
    \label{condition}
\end{equation}

The resolution of the problem is made as follow. For each degree $n$, we compute the coefficients $\alpha_k$  with a least mean square algorithm based on Eq. \ref{EqLMS} and \ref{condition} to have $E$. We then compute $\mid \mid   <P_\text{InSight,Corrected}(t_{Yr_1})> ~-~ \beta<P_\text{InSight,Corrected}(t_{Yr_2})> \mid \mid $ using Eq. \ref{defP}. We iterate on the degree $n$ to find which $E$ is solution of the optimization problem described in Eq. \ref{PbOpt}.

To compute the least-mean square inversion, we use the data acquired at the end of the MY 34, at \Ls > 340\textdegree to remove the effect of local dust storms, and data at \Ls < 21\textdegree (the limit of the MSL dataset that we used). We finally find that the correction $E$ can be written as:

\begin{equation}
    E(T) \approx 5.5273\times 10^{-4} \, T^3 - 0.4284 \, T^2 + 109.6849 \, T   -9.2602\times10^3
\label{corection}
\end{equation}

\begin{figure}[H]
    \centering
    \includegraphics[scale=0.3]{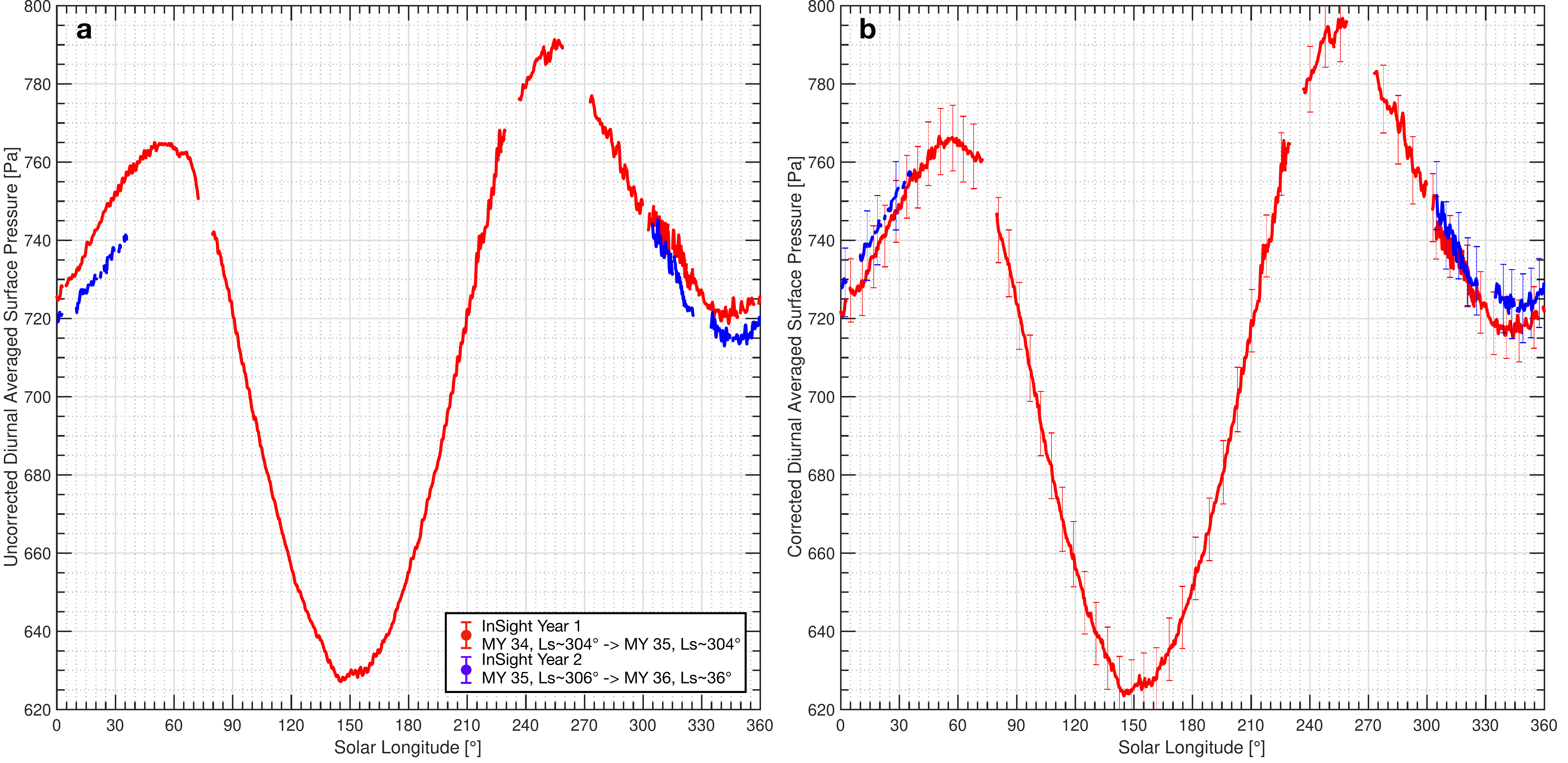}
    \caption{a) Diurnal averaged surface pressure computed from the raw pressure data. b) Diurnal averaged surface pressure after applying the thermal correction. Error bars represent the uncertainty on the measurements after the correction at \sigT. The details of the uncertainty computations are described in the text. Red dots are for the first year or the mission, while blue dots are for the second year.}
    \label{fig:figCor}
\end{figure}

 Applying this correction to the complete measured pressure data with Eq. \ref{defP} leads to the result presented in Figure \ref{fig:figCor}. As expected, this correction strongly modifies the pressure measured by InSight in terms of amplitude and shape. The Northern winter surface pressure during InSight Year 1 is lower than during InSight Year 2, but tends to equalize during spring. These results are thus consistent with the analysis of contemporary MSL data from \citeA{delaTorreJuarez2019}.

\subsection{Uncertainty of the corrected data}
\label{ssec:Sensitivity}

The accuracy of InSight pressure measurements is crucial for the determination of possible secular pressure changes and will be useful to the community for future scientific work. We thus need to quantify the uncertainty of the InSight corrected pressure proposed in section~\ref{ssec:Recalibration}. Three kinds of uncertainty can be highlighted here according to Eq. \ref{defP}. The first one is the uncertainty of the pressure sensor on the measurement, which is $\sigma_{P_{sensor}}$= 1.5 ~Pa~R.M.S   \cite{Banfield2018}. The second one is caused by the uncertainty of the correction. This correction deals with the uncertainties at \sigO of the pressure measured by InSight $\sigma_{P_{sensor}}$ and the sensor temperature uncertainty $\sigma_T$ set here at 1~K. However, the uncertainty of $\beta$, and thus the impact of MSL uncertainty on our correction, is negligible. ($\frac{\sigma_\beta}{\beta} << 1$, Appendix A). Hence, as we use a ratio  of MSL pressure measurements to derive the interannual variability $\beta$ of the InSight pressure data, the absolute MSL pressure accuracy does not impact the accuracy of our correction for the InSight pressure measurements. Finally, an uncertainty is associated with the choice of the temperature nullifying the correction term (Eq. \ref{condition}); this last one having been made arbitrarily after analysis of the correlations between the measured pressure and the sensor temperature.

\begin{figure}[H]
    \centering
    \includegraphics[width = 1.1\textwidth]{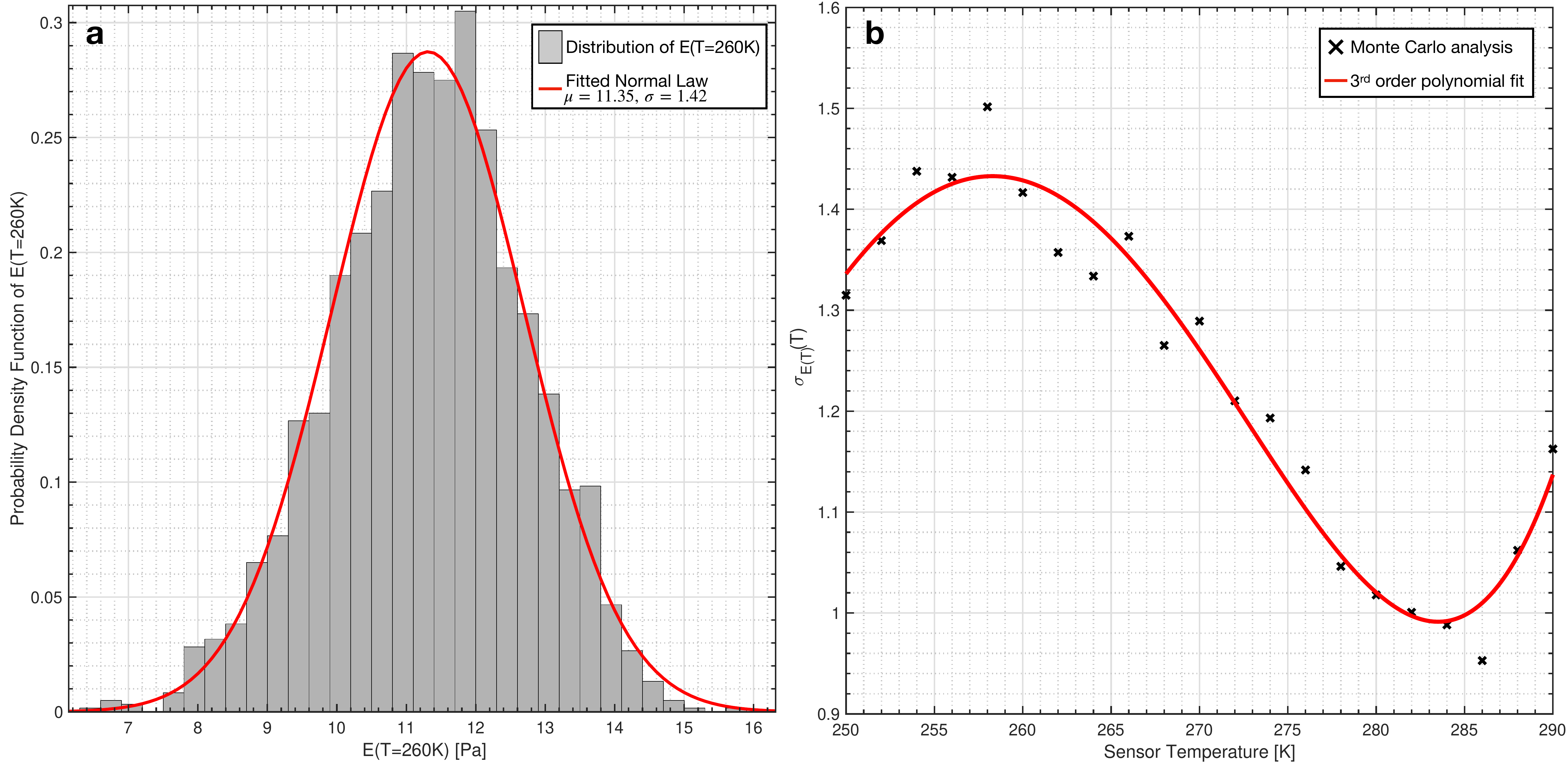}
    \caption{a) Monte Carlo analysis to retrieve  $\sigma_{E(T = 260 K)}$. The histogram of the samples is presented in gray and is normalized to obtain a probability density function.  The fitted normal law is illustrated in red and has as parameters the mean $\mu$  and the standard deviation of the distribution $\sigma$. b) Empirical law for $\sigma_{E(T)}(T)$ obtained from Monte Carlo analysis (black cross) and 3$^{rd}$ order polynomial fit of this law (red line) }
    \label{fig:figUnc}
\end{figure}

To  derive the uncertainty in $E$ at a sensor temperature $T$ ($\Delta E(T))$, we perform a Monte Carlo error analysis as described in \citeA{Press1993} or \citeA{Forget2007}. We generate an ensemble of $10^4$ inputs ($P_\text{InSight}$,$T$), affected by the various uncertainties described above.  All the input parameters are computed using their nominal values plus random values computed from a normal distribution with a standard deviation associated with $\sigma_{P_{sensor}},\sigma_{T}$. The condition provided in Eq. \ref{condition} is also perturbed using a normal distribution with a standard deviation of $\sigma_T$. We then apply our algorithm to retrieve the thermal correction $E$ at a given temperature $T$ with these inputs. We finally compute the standard deviation of the $E$ provided. We find that the distribution of the $E$ retrieved follows a normal distribution as illustrated in Figure \ref{fig:figUnc}a. The standard deviation of the fitted normal distribution gives the uncertainty of $E(T)$ at \sigO.  We apply this Monte Carlo analysis for temperatures ranging from 250~K to 290~K to retrieve $\sigma_{E(T)}$, i.e., the uncertainty at \sigO level. The results from this computation are presented in Figure \ref{fig:figUnc}b. The variations of this curve follow the variations of the gradient of $E(T)$. We then do a least mean square polynomial fit to have an empirical law to simply deduce $\sigma_{E(T)}(T)$:

\begin{equation}
 \sigma_{E(T)}(T) = 5.1453 \times 10^{-5}\,T^3 - 0.0418\,T^2 +   11.2738 \times\,T - 1.0109 \times 10^3
    \label{FitU}
\end{equation}

We finally retrieve the \sigT  uncertainty of one pressure measurements by combining these two uncertainties, plus a term due to the dependence between the measurement and the thermal correction, as the raw measurements and temperature are correlated due to the initial calibration procedure:

\begin{equation}
    \Delta P_{InSight}(T) = 3 \times \sqrt{(\sigma_{P_{sensor}}) ^2 ~+~ (\sigma_{E(T)}(T))^2 ~+~ 2\sigma_{P_{sensor}}\sigma_{E(T)}(T) \rho_{P_{sensor},T }}
    \label{uncertaintyVF}
\end{equation}

\noindent where $\rho_{P_{sensor},T }$ is the correlation coefficient between the raw pressure measurement $P_{sensor}$ and the sensor temperature $(T)$, assumed to be~$1$ as the pressure sensor is calibrated using the sensor temperature \cite{Banfield2018}. Uncertainties on the corrected pressure range from 7.5 Pa to 8.9 Pa  at a \sigT level. Such values are close to the magnitude of the atmospheric mass variations expected based on \citeA{THOMAS2016118}($ \pm$ 9 Pa difference between Viking 1 and InSight surface pressures), but are much smaller than the expected changes that are computed from  \citeA{BLACKBURN2010780,Malin2001}($\sim$+25 Pa  difference between Viking 1 and InSight surface pressure).  We  therefore consider that the InSight corrected pressure data are accurate enough to detect such  secular pressure changes.

\subsection{Comparison with MSL pressure data and validation}
\begin{figure}[h!]
    \centering
    \includegraphics[width = 0.9\textwidth]{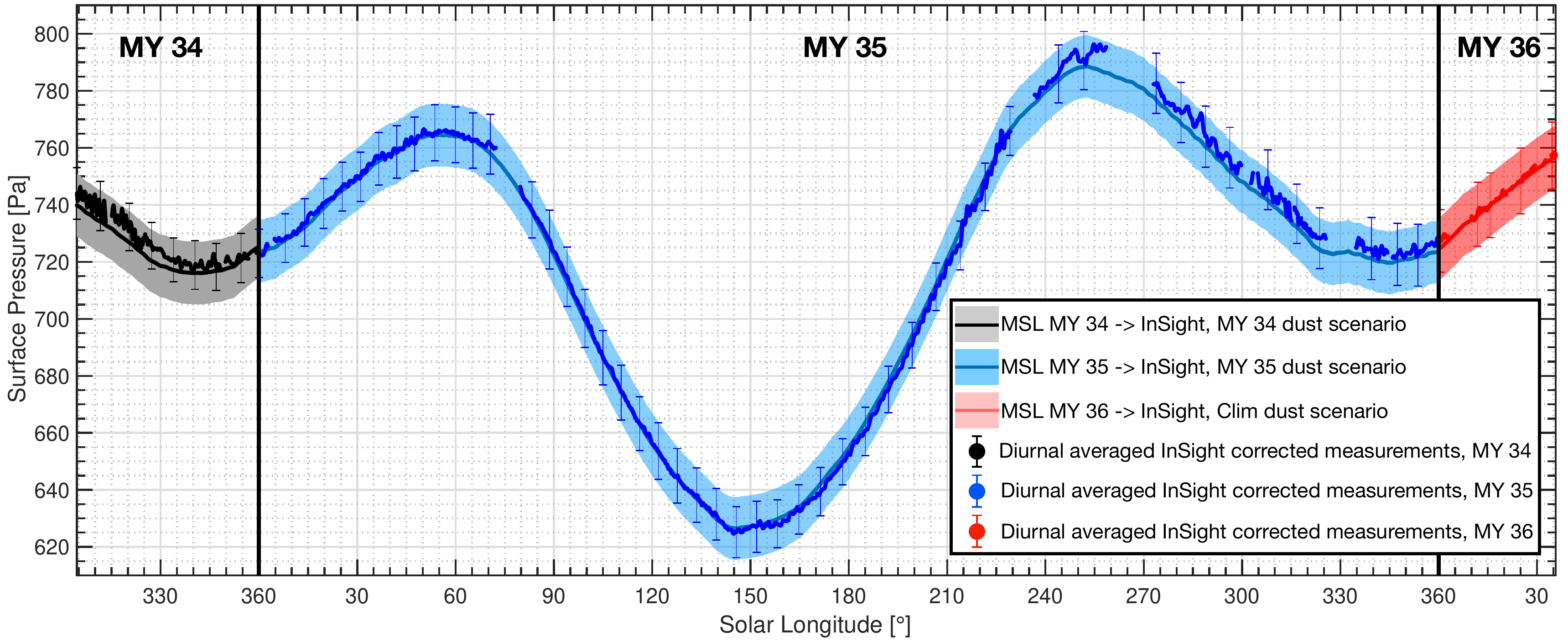}
    \caption{Comparison between the surface pressure measured by InSight and that measured by MSL but interpolated to the InSight landing site, for MY 34, 35, and 36. The filled box around the plain line depicts the \sigT uncertainty of the interpolation due to weather-induced uncertainty and MSL absolute errors, following the methodology presented in section \ref{ssec:GlobalInterp}. Pressure interpolated is averaged over a period of 15 sols to remove atmospheric tides and baroclinic activity. InSight measurements are diurnally averaged thus still indicate baroclinic activity with periods of several sols. The error bars correspond to the \sigT on InSight corrected pressure measurements as described in section \ref{ssec:Sensitivity}.  }
    \label{fig:figCompMSLINS}
\end{figure}

To test the reliability of our correction, we propose here to compare the corrected InSight pressure to the MSL pressure measurements interpolated to the InSight landing site. This comparison is relevant as  the use of the MSL data to correct the InSight data relied on the year-to-year ratio (Eq. \ref{fracbeta}), and thus does not influence seasonal variation given by InSight pressure data after the correction. To do so, we use the methodology described in \ref{ssec:GlobalInterp} by using the ratio  $\frac{P_{MSL}}{P_{GCM,MSL}}$ into Eq. \ref{Interp}, with MY 34,35 and \textit{clim} dust scenario for the beginning of MY 36. The comparison between interpolated MSL pressure and InSight measurements is presented in Figure \ref{fig:figCompMSLINS}. There is an overall good agreement between the InSight corrected measurements and the MSL pressure measurements. This consistency strengthens the credibility of our correction. 

We note a  deficit ~$\delta$ of pressure between MSL pressure interpolated at the InSight landing site, and InSight corrected pressure between \Ls $\sim$ 200\textdegree~and  \Ls < $\sim$ 360\textdegree~(Fig. \ref{fig:figCompMSLINS}). Three causes could explain this deficit: 1) a dynamical effect that is not included in our interpolation process 2) a meteorological effect that changes the thermal state of the atmosphere, and thus the scale height used during the interpolation 3) a problem with our correction.

As InSight and MSL are close to each other ($\sim$ 600 km), the impact of atmospheric dynamics on the interpolation is  limited. As underlined by Figure \ref{fig:figCompMSLINS}, large-scale atmospheric dynamics does not explain the pressure difference.  Another possible explanation might be the small-scale/regional topography of Gale crater that is not included by our GCM. The two closest points are at a longitude of 135\textdegree~E and 140.6\textdegree~E, with an altitude of -2069 and -1879~m respectively. The interpolation using the four closest points of the GCM to the MSL landing site gives an altitude of -1544m, far  from the actual landing site altitude of -4501~m. Furthermore, complex crater circulations \cite[]{PLAGARCIA2016103, RAFKIN2016114} might  impact the pressure measured by MSL. To investigate this deficit, we used the  mesoscale LMD model simulations described in section \ref{ssec:InterpLoc}. We ran the model for 24 hours after initial spin-up time of 24 hours, at \Ls = 270\textdegree~. The mesoscale model, representing a more accurate topography and capturing local circulations, helps to reduce the pressure deficit between MSL interpolated to InSight and InSight measurements but still not fully explain the difference observed. Hence, the deficit~ $\delta$ does not seem to be due entirely to a dynamic effect.

\begin{figure}[H]
    \centering
    \includegraphics[width = 1.1\textwidth]{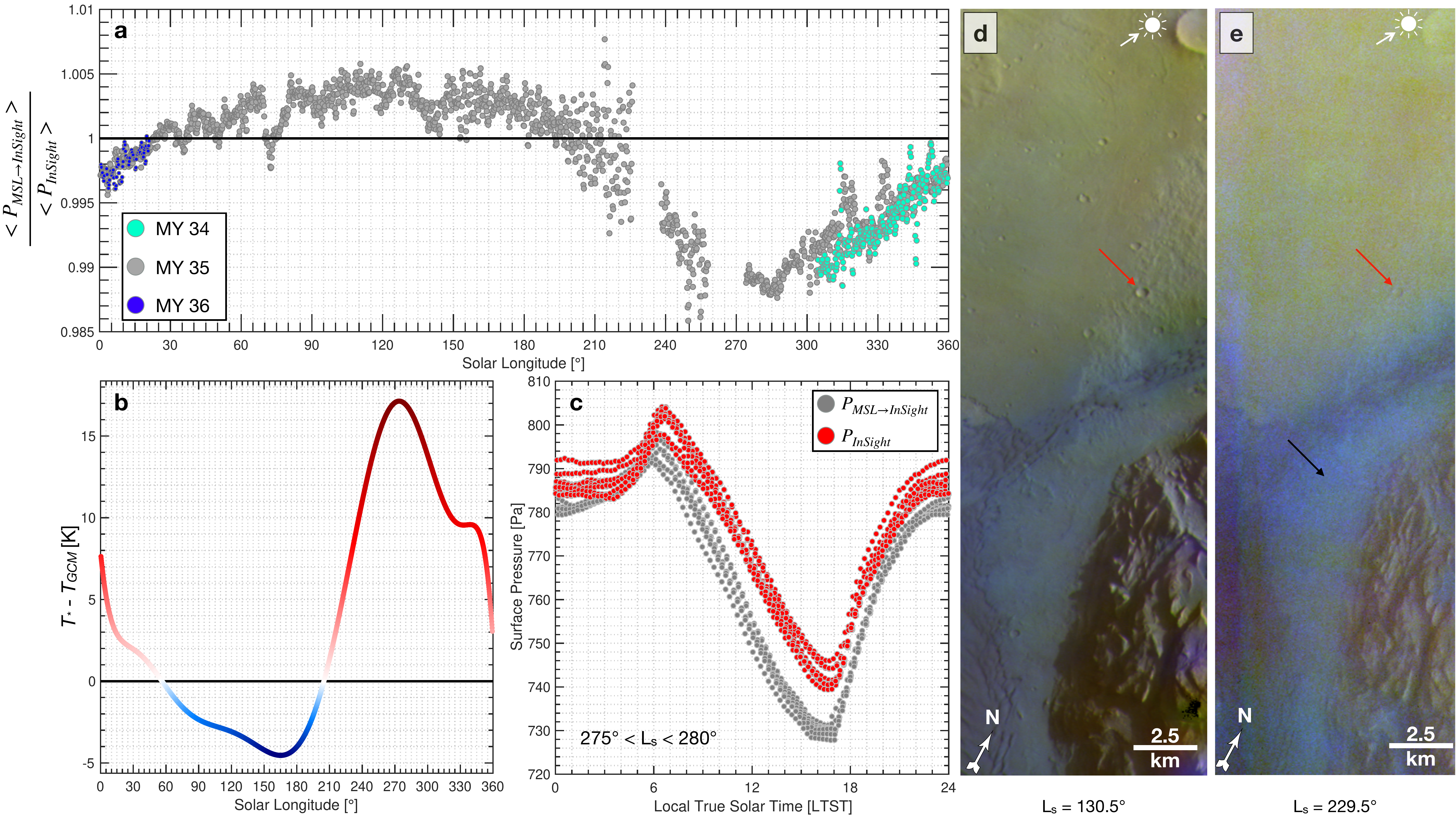}

\caption{a) Evolution of the ratio of MSL REMS pressure measurements interpolated to the InSight landing site, and InSight pressure measurements. Dots correspond to the ratio using the interpolation method described in section \ref{ssec:InterpLoc}, i.e., neglecting atmospheric dynamic effects, during MY 34 (green), MY 35 (grey), and MY 36 (blue). b) Anomaly between the temperature of the GCM at an altitude of 1 km above the surface, and the temperature $T_*$ that gives a ratio of 1, as a function of \Ls (colored curve) for MY~35. c) Comparison between InSight surface pressure over a complete sol and MSL pressure interpolated at InSight landing site between \Ls = 275\textdegree and 280\textdegree, during MY 35. d) Extract of THEMIS image \textit{V63417011} of Gale Crater (center of the original image: 4.9\textdegree S;137.0\textdegree E ) taken at \Ls = 130\textdegree, LTST = 7.2hr, with a solar incident angle of 74.5\textdegree. e) Extract of THEMIS image \textit{V65575024} at the same location, taken at \Ls = 229\textdegree, LTST = 7.2hr, with a solar incident angle of 71.3\textdegree. The black arrow on e) points to the suspected aerosols, whereas the red arrows on d) and e) point to the same crater for a comparison of the perceptibility of the ground. White arrows point to the position of the Sun in the sky.}
    \label{fig:CompINS_MSL_T}
\end{figure}

We then studied this deficit~$\delta$ by investigating the possible influence of the scale height $H$, using the  interpolation  described in \ref{ssec:InterpLoc}. Results are presented in Figure \ref{fig:CompINS_MSL_T}a. We observe again the pressure deficit between InSight and MSL after \Ls = 180\textdegree~. To study the influence of the scale height, we compute the temperature $T_*$ such that: $\frac{<P_{\text{MSL} \rightarrow \text{InSight}}>}{ <P_{\text{InSight}}>} \approx~1$. Using Eq. \ref{Eq:Hydrostatic} and \ref{Eq:H},  $T_*$ writes: 
\begin{equation}
    T_* = -\frac{\Delta z}{\frac{R}{\mu g}\ln{\frac{P_\text{MSL,measured}}{P_\text{InSight,measured}}}}
\end{equation}

\noindent where $\Delta z$ correspond to the difference of altitude between the InSight landing site and MSL altitude (in meters), and $P_{MSL,measured}$ is the raw MSL REMS pressure measurements. 
To detect any anomaly in the temperature, we compare this temperature $T_*$ to the temperature at an altitude of 1~km above the surface predicted by the GCM (Figure \ref{fig:CompINS_MSL_T}b). This comparison underlines warming by 10-15 Kelvin of the temperature at this altitude, at 200\textdegree~ < \Ls < 360\textdegree~.

This difference between $T_*$ and the temperature given by the GCM could be explained by an unexpected accumulation of aerosols within Gale Crater, such as dust, compared to what is assumed in the GCM. The presence of aerosols would indeed  warm up the air as they absorb solar radiation. Moreover, GCM simulations using our \textit{warm} scenario indicate a warming of nearly 5-8~K of this atmospheric layer, which is the order of magnitude of the anomaly observed here.  Hence, by studying the evolution of the temperature anomaly presented in Figure \ref{fig:CompINS_MSL_T}b, we could assume that  at \Ls > 180\textdegree, there are  local effects that increase the quantity of  dust or other aerosols within the crater, inducing  a warming of the air temperature.  By comparing the measured and interpolated pressures at different local times around \Ls = 275\textdegree (Figure \ref{fig:CompINS_MSL_T}c),  we find that this pressure anomaly occurs during the day. We can assume that the air within the crater is heated during the day due to the presence of this dust in suspension. This hypothesis is consistent with the REMS temperature observations at the surface and at 2~m from the ground at this time of year, which are respectively lower and higher than predicted by mesoscale models \cite{PLAGARCIA2016103}.

Several observations reinforce the credibility of this scenario. Measurements of the line-of-sight across-crater extinction with MSL cameras report an increase of dust loading at lower elevations during the dusty season, i.e., 180\textdegree~ < \Ls < 360\textdegree~. The analysis of UV sensors data onboard MSL also confirms this observation, with net dust lifting from the crater floor during the dusty season, and net deposition during the rest of the year \cite{VicenteRetortillo2018}. Such observations  confirmed models of 
dust diffusion rate within Gale crater \cite{MOORE2019197} that report net 
dust lifting from \Ls $\approx$~220-240\textdegree, and net dust deposition in the crater before this date, explaining the variation of the thermal inertia of the ground  \cite{RANGARAJAN2020113499}. This behavior of settling and suspension of dust within Gale might be explained by the dynamics of the planetary boundary layer within the crater. \citeA{FONSECA2018537} points out that at \Ls > 180\textdegree, the planetary boundary layer (PBL) height is higher than the crater rim for a few hours during the afternoon, inducing a mixing between the air outside and inside the crater. In addition, dust might be injected within the crater because of dust devils and wind-driven dust lifting. As reported by \cite{STEAKLEY2016180,Kahanpaa2016,Ordonez2018DROPS,Newman2019},  there are  very strong seasonal variations in dust devil activity, with a peak of  that dust devil lifting  around southern spring and summer, i.e., when we observed the pressure deficit $\delta$. Most of the dust devil occurs during the day, with a peak of activity around noon, close to the period of the sol when the pressure difference between MSL and InSight is the most important (Fig. \ref{fig:CompINS_MSL_T}c).   


We also obtain indications of the presence of  aerosols near the surface using the  THEMIS visible camera \cite{Christensen2004}. Figures \ref{fig:CompINS_MSL_T}d and e compare two images of Gale Crater, at the same local hour, in quasi-similar illumination conditions, but at two different \Ls (130\textdegree~ and 229\textdegree~ respectively). Figure \ref{fig:CompINS_MSL_T}e clearly shows the presence of aerosols (black arrow)  confined within the crater, as the portion of Mount Sharp remains easily detectable and less obstructed. Another indication of the presence of a significant quantity of aerosols in the air is the difficulty of detecting the ground and the craters at the bottom of Gale crater on the image e compared to d (see red arrows), in quasi-similar illumination conditions. Water fog is a suspected candidate to explain this phenomenon, as the image is taken during the early morning, but it seems highly unrealistic as the relative humidity at this time of the year is at its lowest level \cite{Martnez2016, Martnez2017}. Furthermore, images taken at nearly 17~hr Local True Solar Time (LTST) also report such features (see for instance THEMIS image \textit{V59356002} that were taken at a quite similar location at \Ls = 335\textdegree~, at 17.37~hr). However, opacity derived from MSL cameras does not show a significant increase of dust loading at this time, compared to what is predicted by the \textit{clim} scenario  (see Figure 9 of \citeA{ORDONEZETXEBERRIA2019591}).

Hence, even if we have several indications making credible our scenario of dust in suspension that explains the temperature anomaly within Gale Crater,   further investigations are required to confirm this  assumption, including an analysis of UV fluxes for instance. This hypothesis, however, has the potential to explain the observed deficit, and thus does not cast major doubt on our correction to the InSight pressure data. A major mistake in our correction seems unlikely because of the very good agreement between the pressure measurements of MSL and InSight during the rest of the year.

\section{Results: Comparison with Viking Lander 1 Pressure Data\label{sec:Results}}

\begin{figure}[H]
    \centering
    \includegraphics[scale=.5]{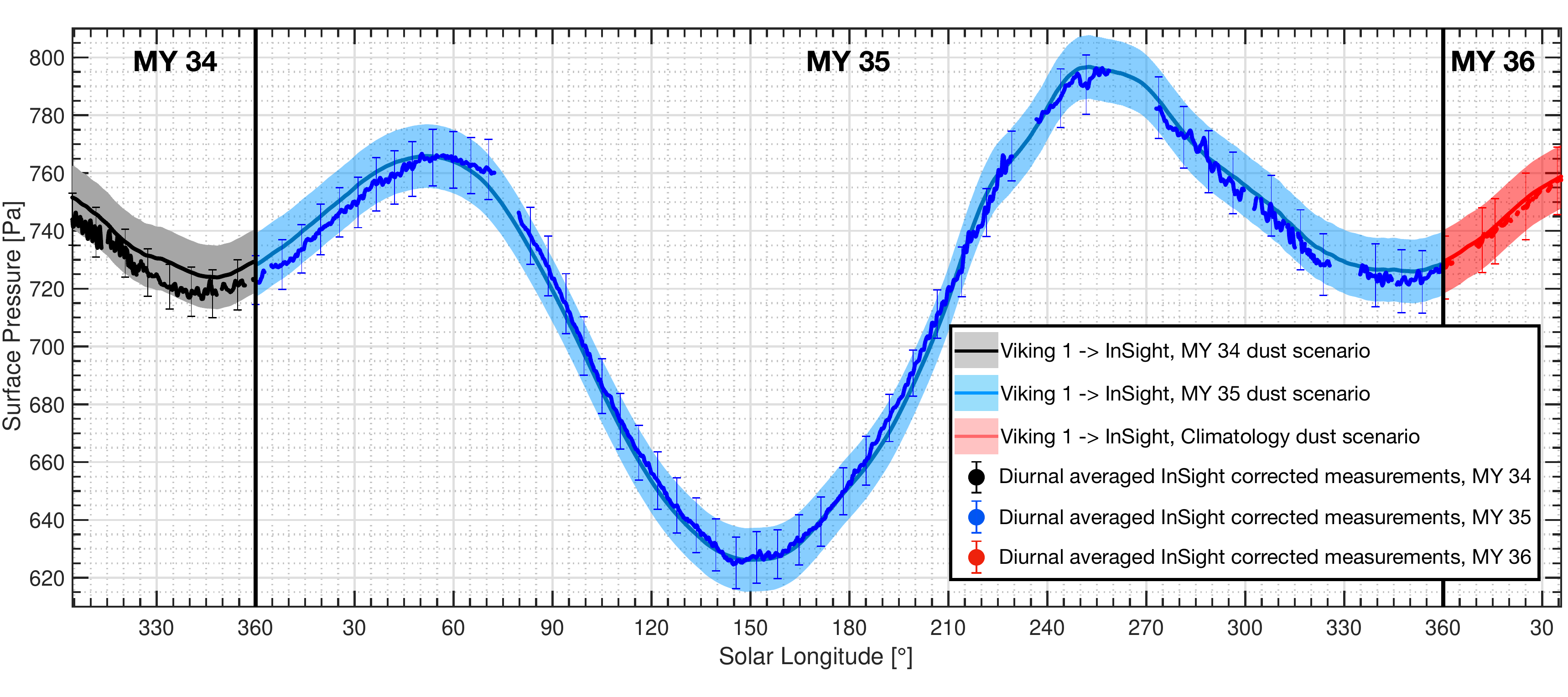}
    \caption{ Comparison between the surface pressure by Viking 1 interpolated at the InSight landing site for MY 34, 35, and 36. The filled box around the plain line depicts the \sigT uncertainty of the interpolation  detailed in section \ref{ssec:GlobalInterp}. Pressure interpolated is averaged on a period of 15\textdegree to remove atmospheric tides and baroclinic activity. InSight measurements are diurnal averaged and still keep baroclinic activity. The error bars correspond to the \sigT on InSight corrected pressure measurements as described in section \ref{ssec:Sensitivity}.}
    \label{fig:figCompVL1INS}
\end{figure}

The comparison between the  Viking 1 surface pressure measurements interpolated at the InSight landing site and the InSight temperature-corrected measurements for MY 34, 35, and 36 is presented in Figure \ref{fig:figCompVL1INS}. During MY 34 and the beginning of MY 35 (\Ls<55\textdegree), InSight pressure measurements are lower compared to Viking 1 pressure by  5-10~Pa. \citeA{delaTorreJuarez2019} also reported a pressure deficit at these times when studying the repeatability of  MSL pressure data.  Using MCS thermal data, they relate this to a possible increase of the NSPC expansion during MY 34 compared to MY 33. Such an expansion would consequently decrease the  atmospheric mass at this time, reduce the surface pressure, and thereby explain the deficit observed. This deficit is not observed during  Northern winter of  MY 35 with MSL and InSight pressure data, and thus cannot be linked to a secular change.

After the sublimation of the NSPC during MY 35, the InSight pressure measurements match Viking pressures very well within the uncertainties associated with the interpolation method. The weather-induced uncertainty might explain the small deficit of pressure observed at \Ls>250\textdegree because Viking 1 pressure was more affected by baroclinic activity as the lander is at a higher latitude than InSight. There is also smaller confidence in the Viking 1 pressure average during this period, as a lot of the measurements available at this time of the year were affected by the first global dust storm recorded by Viking \cite{Ryan1981}, and thus removed from the dataset.

However, the overall strong agreement between Viking 1 interpolated surface pressure and InSight thermally corrected measurements strongly supports the assumption that the atmospheric mass has not changed since the Viking era, nearly forty earth years before the InSight era. Our results suggest that  SPRC mass balances from \citeA{Malin2001, BLACKBURN2010780} might have been overestimated, but support low estimated values of atmospheric mass gain/loss due to the evolution of the SPRC \cite{THOMAS2016118}. Such  results thus reinforce the assumption that the SPRC does not suffer from major changes over decades, as indicated by both imagery comparisons since the Mariner era and recent imagery dataset \cite{Piqueux2008, THOMAS2016118}. In fact, the SPRC might be varying with  periods of erosion due to large summer dust events, followed by a period of deposition in the next winter \cite{Bonev2008,Becerra2015, Byrne2015,Jammes1992Book,James2010, THOMAS2016118}. Further discussion on the role of dust events in the condensation and sublimation of \CO ice is shown  in section  \ref{ss:DustAndVariability}. The possible influence of \CO reservoirs under the SPRC on the durability of this cap has also been explored recently \cite{Buhler2020}.

The strong agreement of the surface pressure comparison during the formation and sublimation of seasonal caps (excluding the Northern winter of MY 34) also suggests a low variability of the martian seasonal CO$_2$ cap dynamic. This result is consistent with those of  \citeA{PIQUEUX2015} using infrared spectroscopy, or those obtained using albedo monitoring \cite{James1979b, James1982, James1982b, CALVIN2015North, CALVIN2017South}. These studies cannot detect changes over decadal timescales, however, as these comparisons are limited to observations separated by a few martian years, or use telescopic observations of the twentieth century that are potentially less accurate \cite{James1987, Piqueux2008}. We propose here to track possible secular change on the seasonality of these caps by making a comparison of the caps' albedo. To do so, we exploit images of the SSPC taken by Viking Orbiter 2  during MY 12 \cite{James1979} and MARCI images \cite{Malin2001MARCI} taken during  MY 35 (images from MY 34 are not used because of the global dust storm that occurred during this year, hindering the visual detection of the caps). \citeA{PIQUEUX2015} noted interannual variability in the caps' dynamics due to global dust storms. We thus add MY 33 to the comparison as a control year in case the global dust storm at the end of MY 34 influenced the cap dynamics during MY 35. Furthermore, even if the cap boundary is composed of water ice  after the sublimation of the seasonal \CO ice (see the spectroscopic study in \citeA{Langevin2007}), we assume that albedo comparison between the Viking decade and late 2010s/early 2020s also reflects possible changes in the \CO cycle.

\begin{figure}[H]
    \centering
    \includegraphics[scale=0.2]{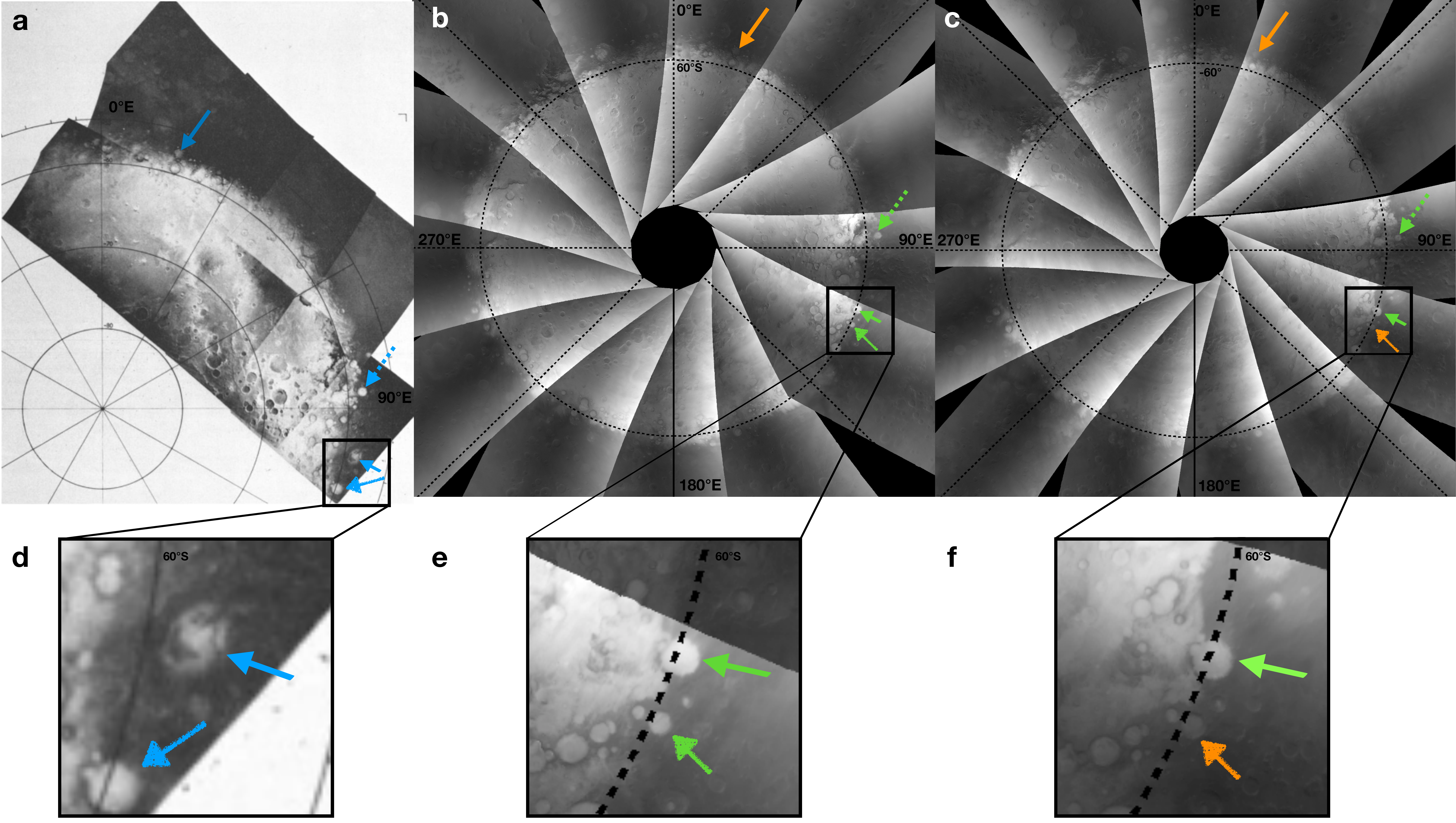}
    \caption{Comparison of the SSPC images taken by (a) Viking orbiter during MY 12, \Ls~= 192.6\textdegree~(extracted from \citeA{James1979}); (b) MARCI during MY 33, \Ls~= 192.3\textdegree~; and (c) MARCI during MY 35, \Ls~= 192.9\textdegree. Blue arrows flag characteristic surface features for the comparison like craters. Orange arrows indicate a possible difference between the Viking images and MARCI images while green arrows indicate a good match between the images. d) to f) are zoom on the lowest flagged craters of a), b), c). The 60\textdegree S circle of latitude on image d) extracted from \citeA{James1979} is misplaced, but arrows point to the same elements.}
    \label{fig:figMarciViking}
\end{figure}

Details on the composition of MARCI polar mosaic are given in \citeA{CALVIN2015North, CALVIN2017South}. We select \Ls = 192.6\textdegree~ for the comparison as Viking mosaic is available at this time (Figure 5 of \citeA{James1979}). Similar analysis and conclusions can be drawn using other \Ls. The comparisons are presented in Figure \ref{fig:figMarciViking}. On Viking images, we flag with blue arrows craters or easily distinguishable topographies that are covered by ice at the boundary of the cap. We then look at MARCI images to see if the element is still covered by ice at the same time of the year. In this case, the element is flagged by a green arrow, whereas in the case of a divergence with Viking observations, the element is flagged by a red arrow.  In case of doubt about the presence of ice, we flag the crater with an orange arrow. The comparisons underline that no major changes have happened in the dynamic of the seasonal caps, thus confirming what has been observed by comparing Viking 1 and InSight pressure data. Little variability can be noted as revealed by the orange arrows. It can be explained by some observational biases. First, MARCI mosaics are built with images taken during all the day, and at different LTST. Hence, some ice might have sublimated during the day and would not be present on the mosaic. Second, the discrepancies on Fig. \ref{fig:figMarciViking}f are actually a consequence of the timing of the mosaic, as the images are not taken exactly at the same \Ls~and illumination conditions. At least, it is very unlikely that these discrepancies observed are due to a faster retreat of the seasonal cap as a consequence of the MY 34 global dust storm. Our GCM simulations show indeed that this retreat should occur at the same speed between MY 33 and 35.

Thus, the comparison between Viking 1 and InSight pressure data, as well as the comparison of images taken by the Viking 2 orbiter and the MARCI camera suggest the absence of secular pressure changes or modifications in the dynamics of the seasonal ice caps.

\section{Discussion\label{sec:Discussion}}

\subsection{Evolution of the atmospheric mass since MY 29}
\label{ssec:Evol_Atm_Mass}

\begin{figure}[H]
    \centering
    \includegraphics[scale = .4]{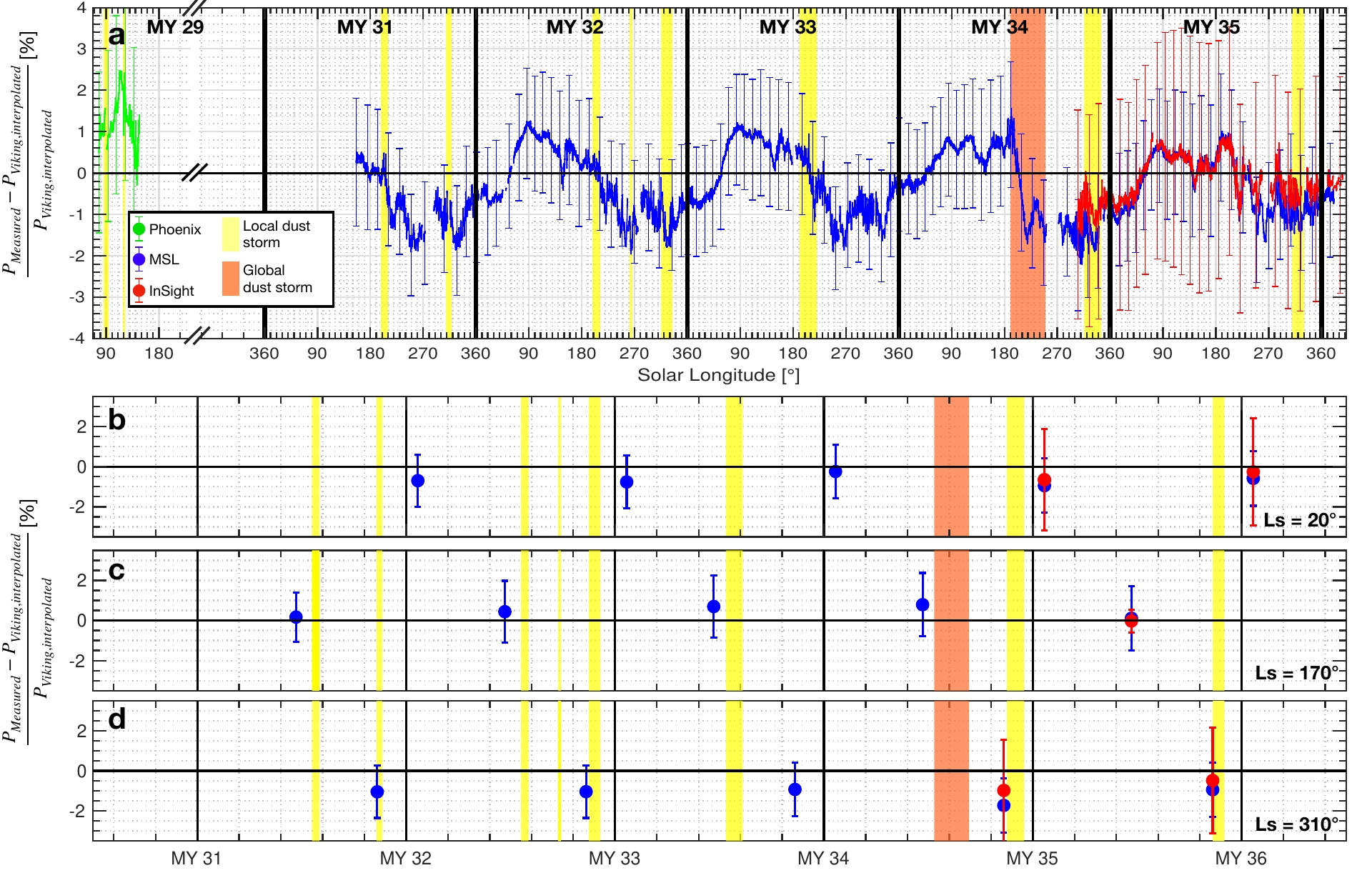}

\caption{a)
Comparison of the surface pressure measured by Phoenix (green dots), MSL (blue dots), and InSight (red dots), to Viking 1 measurements (interpolated at each landing sites), from MY 29 to MY 36.
Yellow boxes correspond to periods of local dust storms at landing sites \cite{Rathlou2010,ORDONEZETXEBERRIA2019591}, while the orange box corresponds to the period of MY 34 global dust storm \cite{Viudez-Moreiras2019}.  b) to d) Evolution of the relative difference between Viking 1 interpolated and MSL (blue) and InSight (red), as a function of Martian Year at \Ls = 20\textdegree (b), \Ls = 170\textdegree (c), \Ls = 310\textdegree (d). The error bars indicate the sensitivity of the comparison with regards to the interpolation uncertainty at \sigT, as described in \ref{ssec:GlobalInterp}. }
    \label{fig:EvolAtmMass_MY29_36}
\end{figure}

The non-detection of atmospheric mass changes between the 1970s and present  disagrees with the conclusions obtained from the comparison between Viking and Phoenix surface pressure made in \cite{Haberle2010}. The preliminary comparison between MSL and Viking 2 pressure data, which are nearly at the same altitude above the surface, did not show any significant increase of the atmospheric mass, but rather possibly a small decrease. We propose to extend this analysis by also comparing Phoenix and MSL data with Viking 1 pressures, using our methodology 

presented in section \ref{ssec:Evol_Atm_Mass} for the comparison between Viking 1 and InSight pressure. Phoenix data used here are extracted from \citeA{Taylor2010}, as they are thus corrected from the temperature gradient within the sensor that disturbed the measurements \cite{Taylor2010}. The results are presented in Figure \ref{fig:EvolAtmMass_MY29_36}.

Figure \ref{fig:EvolAtmMass_MY29_36}a underlines an excess of pressure when comparing Phoenix measurements to Viking 1 interpolated to Phoenix landing site. Such result is consistent with the analysis from \citeA{Haberle2010}. However, the location of Phoenix must be considered and might qualify their conclusions.  Phoenix landed at a high latitude (68.22\textdegree N) compared to the other measurements used in this study that were made at mid/equatorial latitudes. This difference could lead to errors in our interpolation due to the large latitudinal pressure gradients. The error bars in Figure \ref{fig:EvolAtmMass_MY29_36}a underlines that the difference lies within the \sigT uncertainty of our interpolation method. Hence, it is difficult to conclude on a possible increase of atmospheric mass in MY 29 using Phoenix measurements only. These observations might actually illustrate a real rise of the atmospheric mass due to a significant SPRC erosion during the MY 28 global dust storm \cite{MONTABONE201565}. \citeA{ Bonev2008,  Becerra2015,Byrne2015,THOMAS2016118} report that southern spring/summer dust storms, like the one in MY 28, might enhance the SPRC sublimation, which would consequently increase the atmospheric mass. Further observations like an analysis of the SPRC extent between MY 27, 28, 29 should help to find the explanation of this increase of the surface pressure at the Phoenix landing site.

The comparisons of MSL and InSight data with Viking 1 pressure measurements both show the same results, i.e., an excess of pressure for 90\textdegree~< \Ls < 180\textdegree~, and a deficit elsewhere. Such divergences are small (less than 1\% generally) and both comparisons are consistent, i.e., MSL and InSight present a deficit or excess of pressure at the same time of the year, but MSL data have sometimes larger relative differences. We study with Figures \ref{fig:EvolAtmMass_MY29_36}b to d  the evolution of these divergences to Viking interpolated pressure, at three times of the year: \Ls = 20\textdegree, 170\textdegree, 310\textdegree. No clear trend can be established when comparing MSL or InSight to Viking 1 pressure data, thus rejecting the idea of a monotonic SPRC mass balance over the years. The possibility of an erosion of the SPRC following the storm in MY 34 is difficult to show from the pressure data. First, the pressure recorded by MSL at the end of MY 34 was strongly impacted by a local dust storm \cite{VidezMoreiras2020}. Moreover, when the storm stopped, the northern seasonal cap was still forming, with potentially an anomalous extent that lead to a decrease in the available atmospheric mass \cite{delaTorreJuarez2019}.  The uncertainty in both the data and the interpolation, represented by the error bars in Figure \ref{fig:EvolAtmMass_MY29_36}, also explain the deficit observed by InSight and MSL. These two comparisons suggest again that there is no significant long-term pressure change.

\subsection{CO$_2$ cycle and dust }
\label{ss:DustAndVariability} 
What can induce year-to-year variations in the seasonal CO$_2$ ice budget? As reviewed in \citeA{titus_byrne_colaprete_forget_michaels_prettyman_2017}, the CO$_2$ ice condensation and sublimation rates are controlled by the local energy balance, as the CO$_2$
condenses or sublimes in the exact amount needed to keep the surface and atmosphere at the CO$_2$ condensation temperature when ice is present.

At a given season, this energy balance could fluctuate from one year to the other. This stems from interannual changes in both CO$_2$ ice albedo and emissivity, as well as changes in the incident infrared radiation due to variations in the heat advected by the atmosphere or by the clouds. It is also sensitive to the amount of heat stored in the subsurface during previous seasons: the heat conducted from the subsurface up to the CO$_2$ ice on the surface depends on the subsurface temperatures, which are themselves influenced by the temperature from the previous summer when no CO$_2$ ice was present.

On these grounds, atmospheric dust can influence the CO$_2$ budget in a variety of ways:

Firstly, during the condensation phase (i.e. in the polar night), dust primarily increases the thermal emissivity of the atmosphere and thus its radiative cooling \cite{Pollack1990}. More CO$_2$ condenses in the atmosphere and less on the surface. The net effect is an observed decrease of the thermal infrared emission at the top of the atmosphere due to the radiative effect of CO$_2$ clouds and/or the lower emissivity of the CO$_2$ snow freshly deposited from the  atmosphere \cite{Forget1996, Cornwall2009}. This means less CO$_2$ ice condensing during a dust storm reaching the polar night. CO$_2$ ice deposits that condensed in the presence of extra dust may also be durably modified. They could have a higher albedo because they were formed from larger fractions of small particles condensed in the atmosphere, but their albedo could also be lowered by the contamination of more dust particles. Which effects dominate? Looking at the seasonal deposits around the north pole, \citeA{Byrne2008} found that the northern fall 2001 global dust-storm resulted in slightly brighter ice deposits in the following spring. They considered this result to be "counter-intuitive".  It can probably be attributed to comparatively more atmospheric condensation in fall enhancing the spring albedo. The amount of airborne dust  also influences the atmospheric circulation and thus the transport of heat and the dilution of non-condensable gas. The concentration of these gases influences the CO$_2$ condensation temperature \cite{Forget2008,Piqueux2020}.

Secondly, during the sublimation phase, or more generally when CO$_2$ ice is significantly sunlit, the net effect of airborne dust is also equivocal, as studied by \citeA{Bonev2002, Bonev2008}. Airborne dust redistributes the downward radiation from solar to thermal infrared because dust absorbs solar radiations and re-emit at thermal wavelengths. Model calculation and camera observations show that regions of  high-albedo CO$_2$ frost will sublimate faster with more airborne dust (as they mostly absorb in the thermal range)
whereas low-albedo regions will sublimate slower (as they mostly absorb in the visible) \cite{Bonev2003,Bonev2008}.

Thirdly, during summer (when no CO$_2$ ice is present) airborne dust could also modify the mean surface temperature at high latitude and the stored subsurface heat, but once again the net effect is subtle and depends on the atmospheric temperatures and surface albedo.

Overall, determining the net effect of regional and global dust storms on the seasonal CO$_2$ cycle is not straightforward as the different processes involved could tend to balance each other. This may explains why the seasonal cycle was observed to be relatively insensitive to the occurrence or non-occurrence of global dust storms in the multi-year Viking Lander pressure records \cite{Jammes1992Book}.  Now the InSight pressure measurements suggest that the Northern seasonal polar cap was slightly and unusually more massive during the winter and spring of MY~34 (after \Ls=300$^{\circ}$) following an unusual global dust storm that occurred throughout the preceding autumn, well before the observed effect on the seasonal ice cap, in accordance with the observations from MSL \citeA{delaTorreJuarez2019}. Based on the discussion above, we can speculate that the most likely reason for this small excess of mass could be due to a slight increase of the ice albedo, resulting from more atmospheric condensation during fall. An alternative explanation could invoke the fact that the post-storm winter atmosphere in the polar night could be slightly depleted in airborne dust and/or ice clouds compared to regular years, reducing the fraction of CO$_2$ ice clouds and snowfall and therefore increasing the polar night thermal infrared cooling to space,  and thus  the net condensation rate. 

In theory, these hypotheses could be tested using climate simulations performed with a GCM.  The current version of the LMD GCM can account for the effect of dust on the atmospheric dynamics and radiative cooling as
well as their consequence on the atmospheric CO$_2$ condensation and its effect on the polar night emissivity \cite{Forget1998}. However, because of the lack of dust observations in the polar night, the dust climatology available to simulate MY~34 \cite{Montabone2020} in the polar regions is probably not adequate to represent well what happened (either during or after the dust storm). Furthermore, the GCM does not include any feedback on the CO$_2$ ice deposit albedo,  which cannot be affected by the effect of a dust storm (neither the albedo increase due to the additional atmospheric condensation or decrease by the additional dust
contamination).  Nevertheless,  we performed GCM simulations using the MY~34 and MY~35 dust scenarios \cite{MONTABONE201565}, looking for other 
differences that could result from the MY~34 global dust storm. The simulated CO$_2$ mass cycles in the two years were found to be almost almost identical (not shown), confirming that processes that are well represented in the GCM  (e.g. atmospheric dynamic and heat transport, non-condensible gas enrichment) are probably not involved in the interannual seasonal cap variations observed by InSight.

\section{Conclusions\label{sec:conclusion}}
In this study, we compare for the first time the InSight pressure sensor data with Viking data taken 40 years earlier, to detect long-term pressure changes, and with other available surface pressure records. The main conclusions of this investigation are: 

\begin{itemize}
     \item InSight pressure measurements have an unexpected thermal sensitivity to sensor temperature, which dramatically impacts the recorded annual pressure and makes its evolution inconsistent over the two years of the mission.
    \item A polynomial correction in the sensor temperature is proposed, using a ratio of MSL pressure data to account for the interannual variability of the seasonal pressure cycle, observed by MSL between the beginning of MY 34 and 35.
    
    \item InSight data, once recalibrated, have an uncertainty of 1.7 to 2.3 Pa at \sigO compared to the initial uncertainty of 1.5 Pa at \sigO. The correction does not lead to a major uncertainty compromising the detection of secular pressure changes compared to the Viking data, or of interannual changes.
    \item The comparison between MSL and InSight pressure during MY 34 and 35 reinforces the credibility of our correction. This comparison also highlights a pressure deficit at the MSL site at \Ls $\sim$ 270\textdegree. This deficit could be induced by a change in the scale height due to a significant amount of dust within Gale Crater, creating a hot atmospheric layer in the local near-surface atmosphere. 
    
    \item We design two high-accuracy methods for pressure interpolation, at local and global scales, that correct the effects of local and large-scale atmospheric circulations as well as the Martian orography on the seasonal pressure variations. Both methods use a scale height computed with the air temperature at an altitude of 1 km. The influence of atmospheric parameters on this interpolation was quantified at 1\% of the absolute pressure at a \sigT level.
   
    \item The Viking 1 and InSight pressure comparison does not show significant secular pressure change, as previously postulated with the Viking and Phoenix comparison. This suggests that either the sublimation of the SPRC is much slower than expected, or that the system is actually in equilibrium. In any case, it appears that the mass balance computations that predicted a very large increase in atmospheric mass or the rapid SPRC disappearance are overestimated.
    
    \item Similarly, a visual comparative analysis of Viking 2 orbiter and MARCI images of the seasonal ice caps does not show significant change in the dynamics of the seasonal ice caps, as observed when comparing the annual variations of the ice caps with pressure data.
    
    \item Both of these conclusions are also supported by the comparison between MSL and Viking 1 pressure data. Using the five martian years of MSL pressure records, we cannot establish a secular trend. 
    
    \item Phoenix surface pressure data might highlight an increase of the atmospheric mass during MY 29, suggesting a possible erosion of the SPRC after the MY 28 global dust storm. Analysis of the SPRC boundary during MY 27, 28, and 29 would help to study this assumption. 
    
    \item The NSPC is more extended during MY 34 compared to MY 35. However, the physical mechanisms that explain this extent are not understood yet. Investigations conducted with the LMD GCM suggest that atmospheric dynamics, heat transport, or non-condensible gas enrichment are not at the origin of this phenomenon.
    
\end{itemize}

The Perseverance rover that arrived on Mars on February 18$^{th}$, 2021 at a latitude close to Viking 1 lander will provide a unique new pressure dataset to contribute to the study of interannual and secular pressure changes.
Cross-analyses between SPRC evolution, dust storms, and atmospheric mass measurements would also help to better understand the evolution of the SPRC and its relative balance.

    \appendix
\section{Impact of MSL pressure uncertainties on InSight pressure correction\label{app:MSLerror}}

We apply the propagation of uncertainty on the definition of $\beta$ (Eq. \ref{fracbeta}):

\begin{equation}
    \frac{\sigma_{\beta}}{ \beta } =  \sqrt{\left(\frac{\sigma_{P_\text{MSL,Yr 1}}}{P_\text{MSL,Yr 1}}\right)^2 + \left(\frac{\sigma_{P_\text{MSL,Yr 2}}}{P_\text{MSL,Yr 2}}\right)^2 - 2 \frac{Cov({P_\text{MSL,Yr 1},P_\text{MSL,Yr 2}})}{P_\text{MSL,Yr 1}P_\text{MSL,Yr 2}}}
\end{equation}

where $Cov({P_\text{MSL,Yr 1},P_\text{MSL,Yr 2}})$ is the covariance between measurements $P_\text{MSL,Yr 1}$ and $P_\text{MSL,Yr 2}$.

Let us assume that:

\begin{equation}
    P_\text{MSL,Yr 1} = P_\text{MSL}(t_{Yr 1}) = P_\text{atm,true} (t_{Yr 1}) + \epsilon(t_{Yr 1}) + \delta (t_{Yr 1}) 
\end{equation}

\begin{equation}
P_\text{MSL,Yr 2} = P_\text{MSL}(t_{Yr 2})= P_\text{atm,true} (t_{Yr 2}) + \epsilon (t_{Yr 2})+ \delta (t_{Yr 2})
\end{equation}

with
\begin{itemize}
    \item $P_\text{atm, true} (t)$ the true atmospheric pressure that MSL should have recorded without any error 
    \item $\epsilon$(t) the error on a measurement due to:
    \begin{itemize}
        \item The error on the absolute measurement due to the initial calibration, estimated to be at most 4 Pa at 3-$\sigma$ over the possible pressure range at the MSL landing site \cite{Harri2014}. 
        \item The error due to elevation change. During the period considered here, the altitude of the rover  changed by nearly 100~m, which could lead to a change in pressure of 8 Pa at \sigT.
        \item At least, the estimated error is $\sqrt{8^2 + 4^2 } \approx~9$ Pa at 3-$\sigma$
        \item Since 15-day averaged data are used, the uncertainty related to the precision of the measurements is assumed to be negligible.

    \end{itemize}
    \item $\delta(t)$, the drift error which theoretically evolves at a rate of 1 Pa/Martian year at 3-$\sigma$  \cite{Harri2014}.
\end{itemize}

We model these errors by random variables whose variance is given by the previous values. The last two terms in the expression of $P_\text{MSL}$, thus representing the error on the measurement, are random variables of variance. We introduce $\sigma_{P_\text{MSL}} = \sqrt{\sigma_\epsilon^2 + \sigma_\delta^2}$ by independence of these two terms.

Since the errors are computed over the range of possible values of the MSL measurements and not a precise value, the errors $\epsilon, \delta$ are independent of $P_\text{atm,true}$. Moreover, the $P_\text{atm,true}$ between the two years are completely independent. Using the bilinearity of the covariance, and these independences, we obtain:

\begin{equation}
    \label{EqUBeta}
    Cov({P_{\text{MSL},Yr 1},P_{MSL,Yr 2}}) =  Cov(\epsilon (t_{Yr 1}), \epsilon (t_{Yr 2}))  +  Cov(\delta (t_{Yr 1}), \delta (t_{Yr 2}) )
\end{equation}

By definition, for two random variables $a, b$: 

\begin{equation}
    Cov(a,b) = \rho(a,b) \sigma_a \sigma_b
\end{equation}

with $\rho$  the correlation coefficient. $\epsilon$ has been determined during calibration tests and is assumed to be constant over the mission, so that 

\begin{equation}
    Cov(\epsilon (t_{Yr 1}), \epsilon (t_{Yr 2})) = \sigma_\epsilon^2
\end{equation}

\noindent Assuming that the drift grows at a rate of $1$ Pa/MY, we have: 

\begin{equation}
    Cov(\delta (t_{Yr 1}), \delta (t_{Yr 2}) )  = Cov( \delta (t_{Yr 1}), \delta (t_{Yr 1}) + 1  ) = Cov( \delta (t_{Yr 1}), \delta (t_{Yr 1}) )  
\end{equation}

\noindent by property of the covariance. We thus have:
\begin{equation}
    Cov(\delta (t_{Yr 1}), \delta (t_{Yr 2}) )  = \sigma_\delta^2  
\end{equation}

Hence, Eq. \ref{EqUBeta} becomes:

\begin{equation}
    \frac{\sigma_\beta}{\beta} = \sigma_{P_\text{MSL}} \sqrt{\left( \frac{1}{P_{\text{MSL}, Yr 1}} - \frac{1}{P_{\text{MSL}, Yr 2}}\right)^2 }
\end{equation}
    
\noindent which gives:
\begin{equation}
    \frac{\sigma_\beta}{\beta} \sim 5 \times 10^{-5}
\end{equation}

\section*{Open Research}
InSight pressure uncorrected data can be retrieved on the PDS at \url{https://atmos.nmsu.edu/PDS/data/PDS4/InSight/ps_bundle/data_calibrated/} \cite{BanfieldPDS} , MSL REMS data at \url{https://atmos.nmsu.edu/PDS/data/mslrem_1001/DATA/} \cite{MSLPDS}, Viking 1 at \url{https://nssdc.gsfc.nasa.gov/nmc/dataset/display.action?id=PSPA-00526} \cite{TillmanViking}. Phoenix corrected data are given with \citeA{Taylor2010}. Marci mosaics can be reconstructed from the images that are available at  \url{https://pds-imaging.jpl.nasa.gov/data/mro/mars_reconnaissance_orbiter/marci/mrom_0867/data/} for Fig. \ref{fig:figMarciViking}b and \url{https://pds-imaging.jpl.nasa.gov/data/mro/mars_reconnaissance_orbiter/marci/mrom_1197/data/} for Fig. \ref{fig:figMarciViking}c \cite{Malin2001MARCI}. The two THEMIS VIS images are available from the THEMIS Data Node under the Planetary Data System (\url{https://viewer.mars.asu.edu/viewer/themis#P=V65575024&T=2}; \url{https://viewer.mars.asu.edu/viewer/themis#P=V63417011&T=2}) \cite{Christensen2002}. The Mars Climate Database can be retrieved upon request (see \url{http://www-mars.lmd.jussieu.fr/mars/access.html}).

 Data files for figures   used in this analysis are available in a public repository, see \citeA{LangeData}.

\acknowledgments
This project has received funding from the European Research Council (ERC) under the European Union’s Horizon 2020 research and innovation program (grant agreement No 835275). All co-authors acknowledge NASA, the Centre National d’Études Spatiales (CNES) and its partner agencies and institutions, and the flight operations team at JPL, CAB, SISMOC, MSDS, IRIS-DMC, and PDS for providing InSight data. The members of the InSight engineering and operations teams are warmly acknowledged for their dedication and hard work in allowing InSight to perform the numerous measurements used in this study. LL and all French co-authors acknowledge support from the CNES. Part of this work was performed at the Jet Propulsion Laboratory, California Institute of Technology, under a contract with NASA. This study is InSight Contribution Number 248.

%
%

\bibliography{agusample.bib}

%
%
%
%
%

\end{document}